\documentclass[aip,twocolumn,reprint,superscriptaddress]{revtex4-1}
\usepackage{amsmath}
\usepackage{amssymb}
\usepackage[latin9]{inputenc}
\usepackage{graphicx}
\usepackage{hyperref}
\usepackage{siunitx}
\usepackage{graphicx}
\usepackage[color= blue!20]{todonotes}
\usepackage[normalem]{ulem}
\usepackage{color}
\usepackage{ulem}
\usepackage[symbol]{footmisc}

\begin{document}

\onecolumngrid

\noindent\textbf{\textsf{\Large Photon-Pressure Strong-Coupling between two Superconducting Circuits}}

\normalsize
\vspace{.3cm}

\noindent\textsf{D.~Bothner$^*$, I.~C.~Rodrigues$^*$, and G.~A.~Steele}

\vspace{.2cm}
\noindent\textit{Kavli Institute of Nanoscience, Delft University of Technology, PO Box 5046, 2600 GA Delft, The Netherlands\\$^*$\normalfont{these authors contributed equally}}

\vspace{.5cm}

\date{\today}

{\addtolength{\leftskip}{10 mm}
\addtolength{\rightskip}{10 mm}

The nonlinear, parametric coupling between two harmonic oscillators has been used in the field of optomechanics for breakthrough experiments regarding the control and detection of mechanical resonators.
Although this type of interaction is an extremely versatile resource and not limited to coupling light fields to mechanical resonators, there have only been, very few reports of implementing it within other systems so far.
Here, we present a device consisting of two superconducting LC circuits, parametrically coupled to each other by a magnetic flux-tunable photon-pressure interaction.
We observe dynamical backaction between the two circuits, photon-pressure-induced transparency and absorption, and enter the parametric strong-coupling regime, enabling switchable and controllable coherent state transfer between the two modes.
As result of the parametric interaction, we are also able to amplify and observe thermal current fluctuations in a radio-frequency LC circuit close to its quantum ground-state.
Due to the high design flexibility and precision of superconducting circuits and the large single-photon coupling rate, our approach will enable new ways to control and detect radio-frequency photons and allow for experiments in parameter regimes not accessible to other platforms with photon-pressure interaction.
}
\vspace{.5cm}

\twocolumngrid

\section*{Introduction}
\vspace{-2mm}

The nonlinear parametric coupling between two harmonic oscillators has been used in cavity optomechanical systems for groundbreaking developments regarding the detection and control of macroscopic mechanical systems \cite{Aspelmeyer14}.
The most impressive results include the demonstration of ground state cooling of mechanical oscillators with light fields \cite{Teufel11, Chan11}, displacement detection below the standard quantum limit \cite{Teufel09, Anetsberger10}, the quantum-entanglement between distinct mechanical systems \cite{Riedinger18, OckeloenKorppi18} or the generation of non-classical states of motion \cite{Wollman15, Pirkkalainen15, Reed17}.
Besides offering remarkable quantum control over mechanical objects, optomechanical systems have also shown great potential as key elements in quantum information technology, such as quantum-limited parametric microwave amplifiers \cite{Metelmann14, Nunnenkamp14, OckeloenKorppi16, Bothner19}, directional amplifiers \cite{Malz18}, on-chip circulators \cite{Bernier17, Barzanjeh17} or microwave-to-optical frequency transducers \cite{Bochmann13, Andrews14, Forsch19}.
The radiation-pressure type of nonlinear coupling between two harmonic oscillators, basis of all breakthrough experiments and developments in cavity optomechanics, however, is not limited to mechanical systems coupled to light fields.
It can, in principle, be implemented for any system in which the amplitude of one oscillator couples to the resonance frequency of a second as demonstrated in purely mechanical multi-mode systems \cite{Mahboob12, DeAlba16, Mathew16, Cho18}.
Recently, photon-pressure coupling has also been theoretically discussed \cite{Johansson14, Kim15, Hardal17} and realized in a first experiment \cite{Eichler18} between two superconducting circuits.
This approach, which can be considered as an analogue of general optomechanical-like systems utilizing circuits, provides a rich variety of possibilities regarding novel devices and experiments with superconducting quantum circuits.
Due to the large design flexibility and precision in resonance frequency and quality factor of superconducting circuits, experiments in unconventional parameter regimes such as the reversed dissipation \cite{Nunnenkamp14} and the reversed resonance frequency regimes \cite{Jansen19} or the optomechanical single-photon strong coupling regime \cite{Nunnenkamp11, Rabl11} are at reach.
In addition, this scheme allows for the realisation of many recently developed optomechanical quantum technologies based on circuits only, which is highly compatible with the rapidly developing field of superconducting quantum processors and avoids the difficulties related to vibrational system noise and mechanical oscillator reproducibility.
Photon-pressure coupling between two superconducting circuits has also been discussed as novel approach for quantum information processing with Gottesmann-Kitaev-Preskill (GKP) qubits \cite{Gottesmann01, Weigand19}.
Finally, it provides the potential for totally new ways to manipulate, cool and detect radio-frequency photons \cite{Gely19}, relevant for fields ranging from radio-astronomy to nuclear magnetic resonance imaging.
\begin{figure*}
\centerline{\includegraphics[trim = {0cm, 0cm, 0cm, 0cm}, clip=True, width=0.99\textwidth]{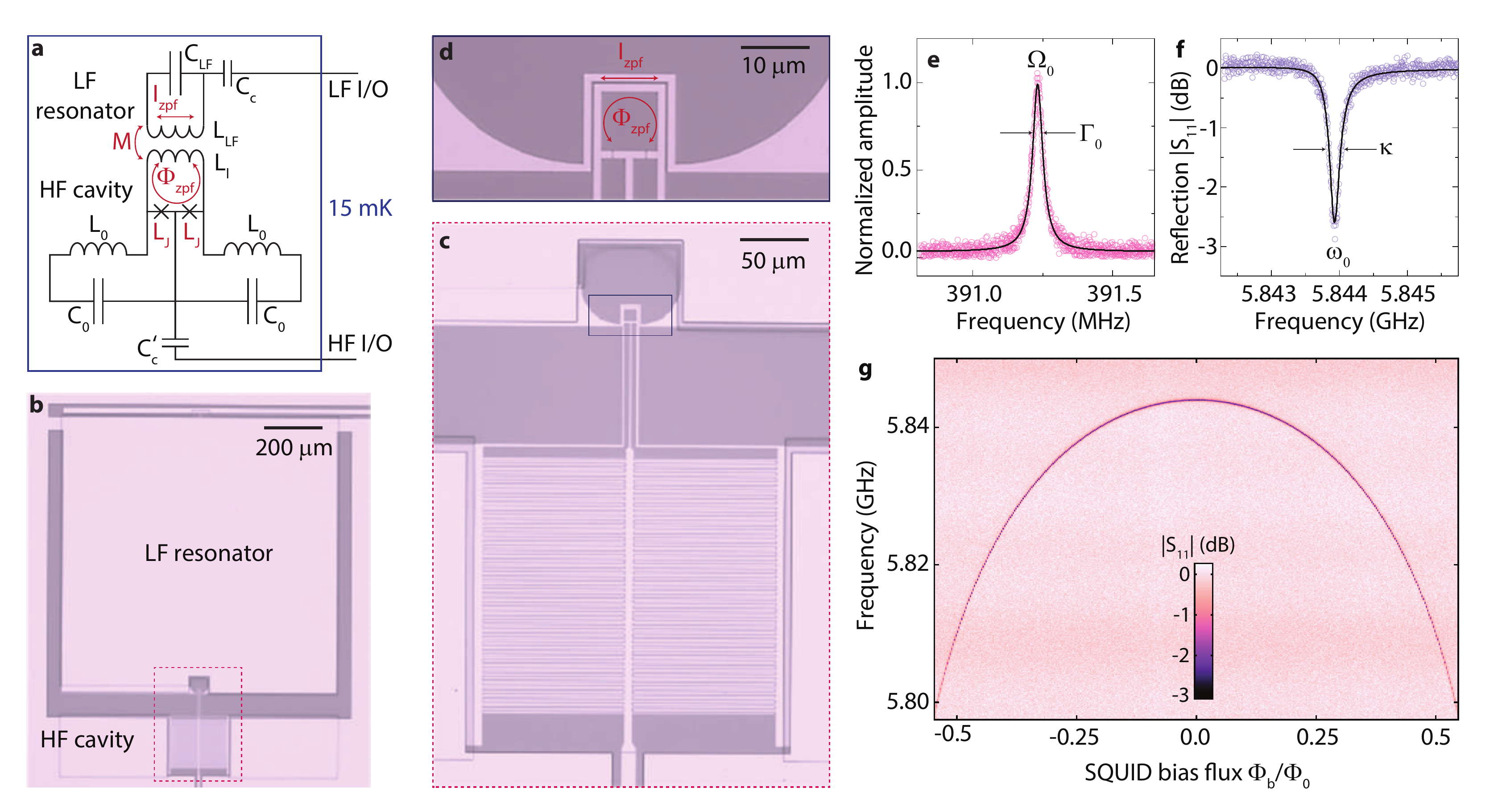}}
\caption{\textsf{\textbf{Two superconducting LC circuits coupled by a photon-pressure interaction.} \textbf{a} Circuit equivalent of the device. The current of a radio-frequency LC circuit is coupled via mutual inductance to a superconducting quantum interference device embedded into a microwave cavity. Both circuits are capacitively coupled to individual feedlines for driving and readout. \textbf{b} Optical image of the device showing both circuits. The red dashed box shows the zoom region for panel \textbf{c}. Panel \textbf{c} shows the high-frequency SQUID cavity with feedline at the bottom, interdigitated capacitors to ground in the center, and linear inductance wires grounding the SQUID symmetrically on both sides. The black box indicates the zoom region for panel \textbf{d}, which shows in detail the SQUID loop and the LF inductance wire which surrounds the loop with a gap of $500\,$nm. In \textbf{b}-\textbf{d}, brighter parts correspond to Aluminum, darker and transparent parts to Silicon. \textbf{e} shows the resonance curve of the LF resonator vs excitation frequency, measured by coherently exciting the LF resonator and using the driven HF SQUID cavity as interferometer. \textbf{f} shows the amplitude of the reflection coefficient $|S_{11}|$ at the SQUID cavity vs excitation frequency. In \textbf{e} and \textbf{f}, colored points are data, and the black lines correspond to fits. In \textbf{g} we show the reflection response of the SQUID cavity dependent on the external magnetic bias flux. For increasing flux, the resonance frequency of the cavity absorption (dark line) is shifted towards lower values. The magnetic field was applied by a small coil below the device, mounted inside the cryoperm magnetic shields surrounding the sample.}}
\label{fig:Device}
\end{figure*}
Here, we present a device consisting of two superconducting microwave resonators, which are coupled to each other by a photon-presssure interaction with a considerable single-photon coupling strength of about 10 percent of the largest system decay rate.
One resonator is a radio-frequency circuit with a resonance frequency in the MHz regime and the second is a microwave quantum interference cavity in the GHz regime.
We demonstrate dynamical backaction between the two LC circuits and observe the transition from the photon-pressure-induced transparency (PPIT) \cite{Agarwal10, Weis10} to the parametric strong-coupling regime \cite{Teufel11a}, manifested by the observation of a pronounced normal-mode splitting.
Finally, we interferometrically observe the photon-pressure-amplified thermal current fluctuations in a radio-frequency LC oscillator by a blue-detuned sideband pump tone.

\section*{Results}
\vspace{-2mm}

\subsection*{Concept and device}
\vspace{-2mm}

Our device combines two superconducting LC circuits with about an order of magnitude difference in resonance frequency.
The full circuit schematic and optical images of the device are shown in Fig.~\ref{fig:Device}\textbf{a}-\textbf{d}.
The low-frequency (LF) resonator consists of a large parallel plate capacitor, whose plates are connected via a short inductor wire, and it is capacitively coupled to a coplanar waveguide feedline for driving and readout.
It has a resonance frequency $\Omega_0 = 2\pi\cdot391\,$MHz and a linewidth $\Gamma_0 = 2\pi\cdot 22\,$kHz, cf. Fig~\ref{fig:Device}\textbf{e}.
The inductor wire of the LF resonator surrounds a superconducting quantum interference device (SQUID) in close proximity, which is embedded into the inductance of a high-frequency (HF) cavity with resonance frequency $\omega_0 = 2\pi\cdot5.844\,$GHz and linewidth $\kappa = 2\pi\cdot250\,$kHz, see Fig.~\ref{fig:Device}\textbf{f}.
The HF SQUID cavity is formed by two interdigitated capacitors and two linear inductors, which are connected to the SQUID loop in the center of the cavity, and it is capacitively coupled to a coplanar waveguide feedline for driving and readout.
Both, inductance and resonance frequency of the HF SQUID cavity depend on the magnetic flux threading the SQUID loop $\omega_0(\Phi) = 1/\sqrt{L_\mathrm{HF}(\Phi)C_\mathrm{HF}}$ as shown in Fig.~\ref{fig:Device}\textbf{g} and can be tuned by applying a magnetic field perpendicular to the chip surface, in our case generated by an external coil below the chip.
The device is mounted to the mixing chamber of a dilution refrigerator with a base temperature $T_b = 15\,$mK, details on the setup can be found in the Supplementary Material (SM) Sec.~S2.
All details regarding the device fabrication, device parameters and modelling are described in the SM Secs.~S1 and S3-S5.
A radio-frequency current flowing through the inductor wire of the LF resonator will couple oscillating flux into the SQUID of the HF cavity and thereby modulate its resonance frequency, giving rise to a parametric photon-pressure interaction between the two circuits.
With the creation and annihilation operators $\hat{a}^\dagger, \hat{a}$ and $\hat{b}^\dagger, \hat{b}$ for the SQUID cavity and LF resonator, respectively, the Hamiltonian of the system is given by
\begin{equation}
\hat{H} = \hbar\omega_0\hat{a}^\dagger\hat{a} + \hbar\Omega_0\hat{b}^\dagger\hat{b} + \hbar g_0\hat{a}^\dagger\hat{a}\left(\hat{b} + \hat{b}^\dagger\right).
\end{equation}
The single-photon coupling rate in the interaction part of the Hamiltonian is given by
\begin{equation}
g_0 = \frac{\partial\omega_0}{\partial\Phi}\Phi_\mathrm{zpf}
\end{equation}
with the SQUID cavity flux responsivity $\partial\omega_0/\partial\Phi$ and the zero-point flux fluctuations $\Phi_\mathrm{zpf} = MI_\mathrm{zpf}$ of the LF resonator.
The zero-point fluctuations of the current are given by $I_\mathrm{zpf} = \sqrt{\frac{\hbar\Omega_0}{2L_\mathrm{LF}}} \approx 21\,$nA, which with the mutual inductance $M = 14\,$pH translates to zero-point flux fluctuations threading the SQUID loop of $\Phi_\mathrm{zpf} = 145\,\mu\Phi_0$.
The flux responsivity $\partial\omega_0/\partial\Phi$ is determined by the flux biasing point, cf. Fig.~\ref{fig:Device}\textbf{g}, and can be tuned in-situ by changing the external SQUID flux bias.
We model the SQUID cavity here as a harmonic oscillator without Kerr-nonlinearity, which is justified by its relatively small anharmonicity.
Due to the small Josephson inductance of the used constriction type Josephson junctions $L_\mathrm{J} = \Phi_0/2\pi I_c \sim 30\,$pH and the inductance dilution of $L_\mathrm{J}/L_\mathrm{tot} \sim 0.04$, the measured frequency shift per photon at the sweetspot $\chi \sim 2\pi\cdot 3\,$kHz (cf. SM Sec.~S4) is much smaller than the cavity linewidth and $\chi/\kappa \sim 10^{-2}$.
In the driven multi-photon regime, the system is well described by the linearized interaction Hamiltonian
\begin{equation}
\hat{H}_\mathrm{int} = \hbar g(\delta\hat{a} + \delta\hat{a}^\dagger)(\hat{b} + \hat{b}^\dagger)
\end{equation}
with the multi-photon coupling rate $g = \sqrt{n_c}g_0$, the field fluctuation creation and annihilation operators $\delta\hat{a}$ and $\delta\hat{a}^\dagger$ and the equilibrium intracavity photon number $n_c$ of the SQUID cavity.
We note here that the SQUID cavity resonance frequency $\omega_0$ and linewidth $\kappa$ depend on both, the flux bias point as well as the intracavity photon number, while the flux bias point also impacts the anharmonicity $\chi$.
Additional data and a detailed discussion can be found in the SM Sec.~S4.

\subsection*{Dynamical backaction between two circuits}

A famous consequence of the parametric photon-pressure interaction is the possibility to manipulate the quality factor and the resonance frequency of the low-frequency resonator by applying a strong coherent pump tone to the high frequency cavity around one of its sidebands $\omega = \omega_0 \pm \Omega_0$.
This effect, which corresponds to a modification of the LF resonator susceptibility, is known in optomechanical systems as dynamical backaction \cite{Schliesser06, Teufel08}.
It arises from a retarded adjustment of the HF intracavity fields and therefore of the photon-pressure coupling to changes of the SQUID cavity resonance frequency induced by the LF flux threading the SQUID loop.
\begin{figure}
\centerline{\includegraphics[trim = {1.0cm, 3.5cm, 0.0cm, 0.0cm}, clip=True,scale=0.55]{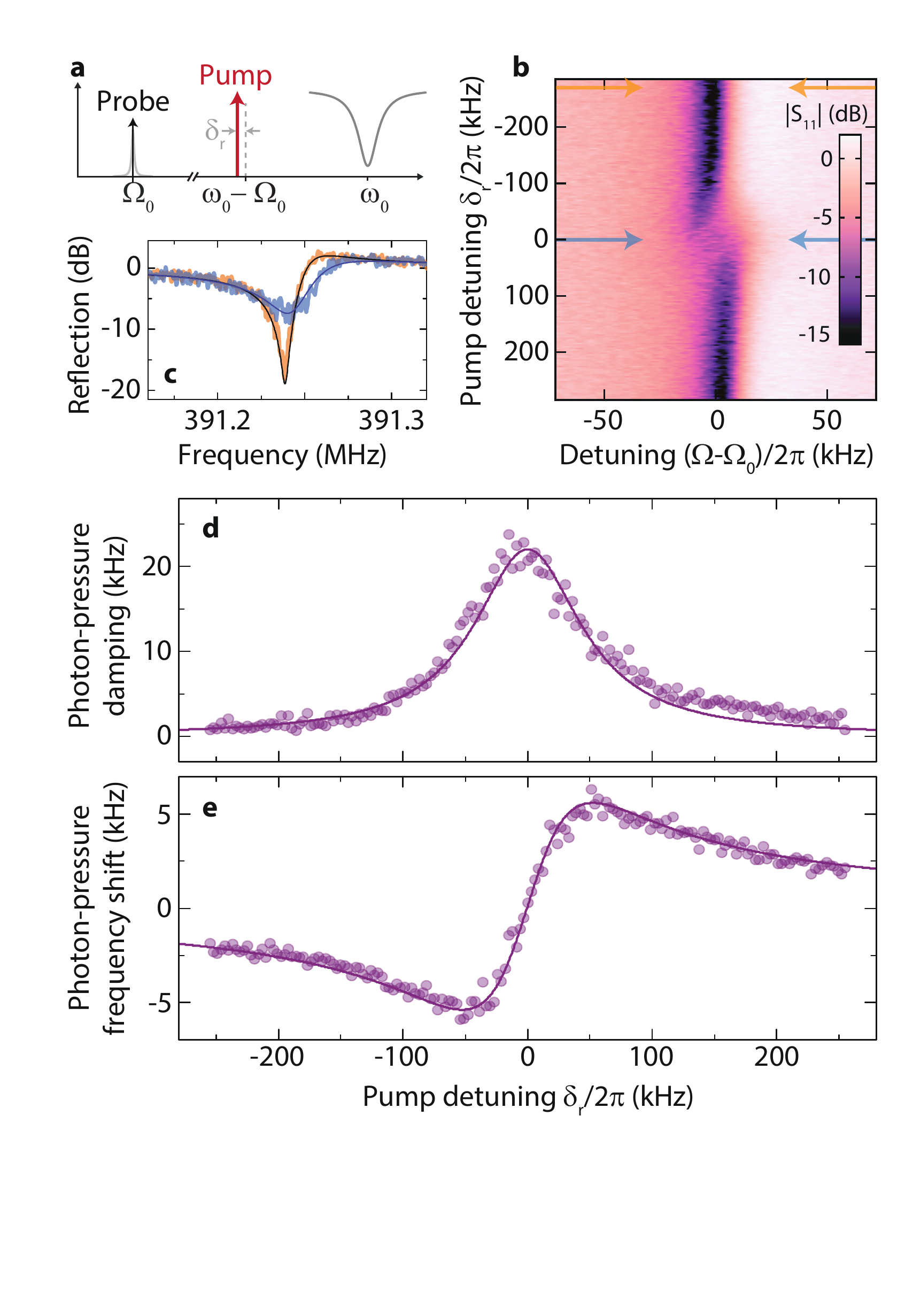}}
\caption{\textsf{\textbf{Observation of photon-pressure dynamical backaction between two superconducting circuits.} \textbf{a} Schematic of the measurement scheme. After flux biasing the cavity to $\Phi = 0.14 \Phi_0$, a pump tone with $\omega_\textrm{p} = \omega_0 - \Omega_0 + \delta_r$ is swept stepwise through the red sideband of the high frequency SQUID cavity. For each pump detuning, the reflection at the low-frequency resonator was scanned with a weak radio-frequency probe tone $\Omega \sim \Omega_0$ and the corresponding reflection parameter $S_{11}$ was measured. The result is shown color-coded in \textbf{b}, where a strong modification of the resonance is visible around $|\delta_r| \lesssim \kappa/2$. In \textbf{c}, a linescan for a far detuned pump $\delta_r = 2\pi\cdot 260\,$ kHz and another for a pump tone exactly on the red sideband $\delta_r = 0$ are shown as orange and light blue curves, respectively. Two pairs of arrows in \textbf{b} visualize the linescan positions. From the corresponding fit curves, shown as black lines in \textbf{c}, the pump-detuning dependent LF resonance frequency $\Omega_0'$ and linewidth $\Gamma_0'$ were extracted. This fitting procedure was repeated for each $\delta_r$. The extracted addition to the low-frequency linewidth, the photon-pressure damping $\delta\Gamma_0 = \Gamma_0' - \Gamma_0$, and the photon-pressure induced shift of the LF resonance frequency $\delta\Omega_0 = \Omega_0' - \Omega_0$ are plotted in \textbf{d} and \textbf{e} as circles, respectively. The lines are theoretical curves, for details see main text.}}
\label{fig:Backaction}
\end{figure}
For the observation of dynamical backaction between the two circuits, we iteratively sweep a pump tone with $\omega_\mathrm{p} = \omega_0 + \Delta$ through the red sideband of the high frequency cavity, i.e., $\Delta = -\Omega_0 + \delta_r$, for details on the experimental setup, cf. SM Sec.~S2.
The cavity is flux biased here at $\Phi = 0.14 \Phi_0$.
For each value of $\delta_r$, we measure the reflection response of the LF resonator $S_{11}$ by probing it directly with a weak radio-frequency probe tone $\Omega \sim \Omega_0$, cf. the schematic in Fig.~\ref{fig:Backaction}\textbf{a} and the response shown in \textbf{b} and \textbf{c}.
In the regime $|\delta_r| \lesssim \kappa/2 $, the low-frequency resonance absorption dip experiences a significant modification in shape, shifting both in resonance frequency and linewidth.
For each pump detuning, we extract resonance frequency $\Omega_0'$ and linewidth $\Gamma_0'$ of the LF resonator from a fit to its measured response (cf. SM Sec.~S3) and determine the photon-pressure induced contributions by substracting the intrinsic values $\Omega_0$ and $\Gamma_0$. 
The resulting frequency shift $\delta\Omega_0 = \Omega_0' - \Omega_0$ and photon-pressure damping $\delta\Gamma_0 = \Gamma_0' - \Gamma_0$, known in optomechanical systems as optical spring and optical damping, respectively, are plotted in Figs.~\ref{fig:Backaction}\textbf{d} and \textbf{e}.
The increase of linewidth for $\delta_r = 0$ is about $\delta\Gamma_0 = 2\pi\cdot22\,$kHz $\sim \Gamma_0$, i.e., of the same magnitude as the intrinsic damping rate, indicating a cooperativity $\mathcal{C} = \frac{4g^2}{\kappa\Gamma_0} \sim 1$ for the chosen parameters.
The lines in Fig.~\ref{fig:Backaction}\textbf{d} and \textbf{e} are simultaneously adjusted theoretical curves using the expressions for photon-pressure induced dynamical backaction in the resolved sideband regime, $\Omega_0 \gg \kappa$,
\begin{eqnarray}
\delta\Omega_0 & = & 4g^2\frac{\delta_r}{\tilde{\kappa}^2 + 4\delta_r^2}\\
\label{eq:spring}
\delta\Gamma_0 & = & 4g^2\frac{\tilde{\kappa}}{\tilde{\kappa}^2 + 4\delta_r^2}
\label{eq:damping}
\end{eqnarray}
and give an excellent agreement with the experimental data for $\tilde{\kappa} = 2\pi\cdot 110\,$kHz, indicating that we indeed observe dynamical backaction between two superconducting circuits.
The reduced cavity linewidth $\tilde{\kappa}$ compared to the directly probed resonance shown in Fig.~\ref{fig:Device} originates from two effects.
First, the SQUID cavity linewidth is power dependent, cf. SM Sec.~S4.
And secondly is the effective SQUID cavity linewidth reduced by the onset of mode hybridization close to the strong-coupling regime.

\begin{figure*}
\centerline{\includegraphics[trim = {3.0cm, 9.5cm, 5.0cm, 0.0cm}, clip=True, width=0.9\textwidth]{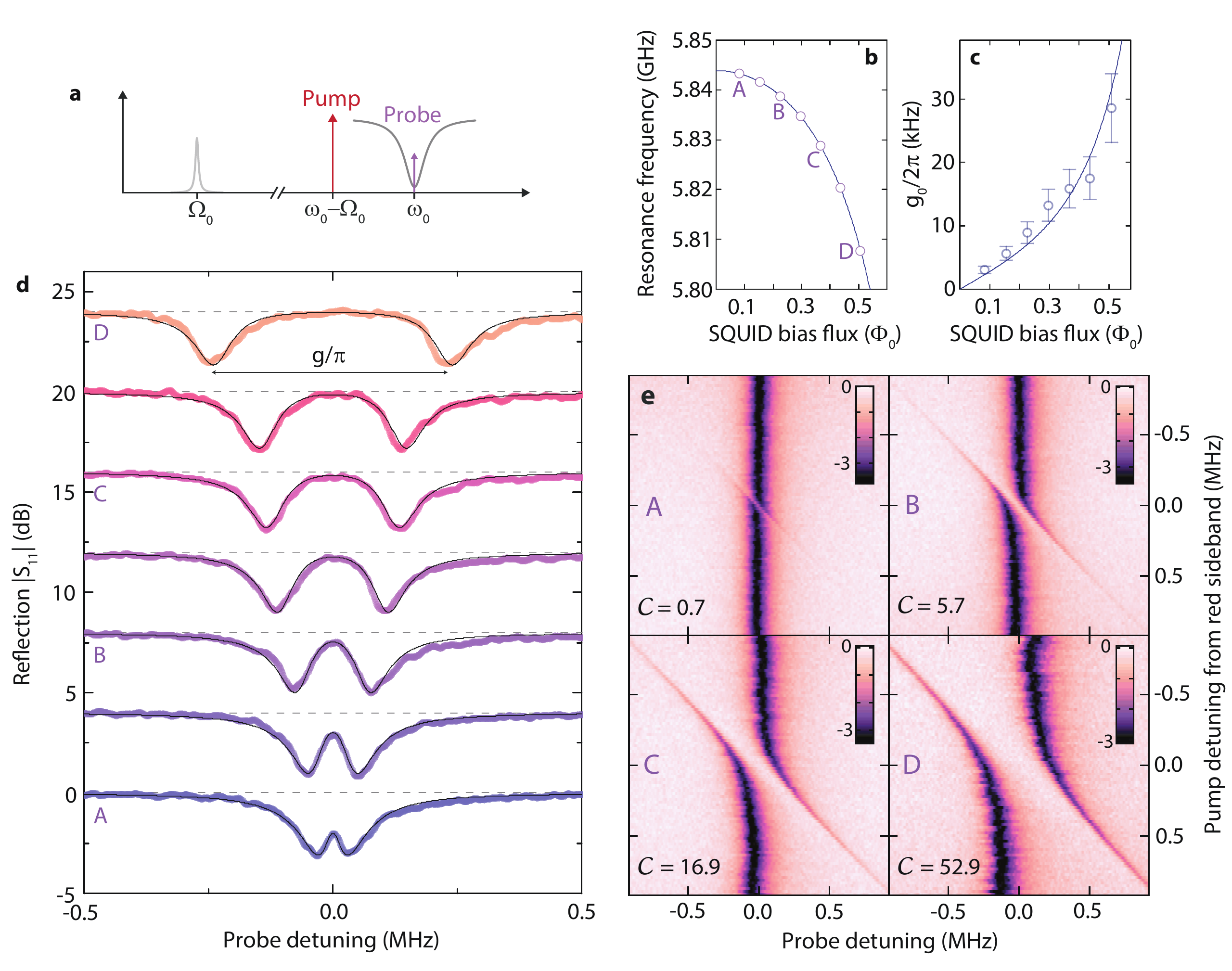}}
\caption{\textsf{\textbf{From photon-pressure induced transparency to the parametric strong-coupling regime by tuning the SQUID flux bias.} \textbf{a} Schematic of a photon-pressure-induced transparency (PPIT) experiment. A pump tone is set to the red sideband of the SQUID cavity $\omega_\textrm{p} = \omega_0 - \Omega_0$, while a weak probe tone is scanning the SQUID cavity response around resonance with $\omega_\textrm{pr} \sim \omega_0$. \textbf{b} shows the SQUID cavity resonance frequency vs magnetic flux as line, together with the seven flux bias points used for the PPIT experiment here, indicated as circles. By increasing the bias flux through the SQUID, the cavity flux responsivity $\partial\omega_0/\partial\Phi$ and therefore the single-photon coupling strength $g_0$ are increased accordingly. This can be seen in panel \textbf{c}, where the expected single-photon coupling rate $g_0$ vs flux bias is shown as line. The measurement configuration described in \textbf{a} was performed for the seven different flux-bias points shown in \textbf{b}, and for each of these biasing points the SQUID cavity reflection $|S_{11}|$ is shown in \textbf{d}. From bottom to top, the flux bias value is increased, and subsequent data are shifted by $+4\,$dB for better visibility. For the lowest flux bias value, we find a small peak in the center of the SQUID cavity absorption dip, indicating the PPIT regime. For larger flux bias values, the PPIT window grows in both, amplitude and width, ultimately leading to two distinct absorption resonances for the largest flux value (top curve). In this regime, where the frequency splitting between the two modes is given by $g/\pi \approx 500\,$kHz, the system has entered the parametric, photon-pressure induced strong-coupling regime. The values for $g_0$ extracted from the theoretical black lines added to the data in \textbf{d}, are plotted as circles in \textbf{c}. In \textbf{e}, we show the SQUID cavity reflection $|S_{11}|$ (color scale given in dB) for four distinct flux bias points, denoted in \textbf{b} with A, B, C, and D, and for non-zero pump detunings $\delta_r \leq 2\pi\cdot 0.9\,$MHz. For small flux bias (A), a small transparency signature is slicing through the cavity response. With increasing flux bias, this transparency window gets stronger and wider, developing into a pronounced normal-mode splitting for the largest flux bias value (D). The intracavity photon number for all data shown here was $n_c \approx 70$ and the cooperativities $C$ for the biasing points A-D are given within the sub-panels of \textbf{e}.}}
\label{fig:strongcoupling}
\end{figure*}
More data on dynamical backaction for different pump powers as well as for a blue-detuned pump frequency with $\Delta\approx +\Omega_0$ can be found in the SM Sec.~S6.

\subsection*{From photon-pressure induced transparency to the parametric strong-coupling regime}

When we choose a similar experimental setting as before, but probe the high-frequency SQUID cavity instead of the radio-frequency circuit with a second weak microwave tone, cf. Fig.~\ref{fig:strongcoupling}\textbf{a} and SM Sec.~S2, a new effect occurs, which we call photon-pressure induced transparency (PPIT), similar to electromagnetically induced transparency in atoms \cite{Fleischhauer05} and optomechanically induced transparency in optomechanical systems \cite{Agarwal10, Weis10}.
A pump tone on the red cavity sideband with $\omega_\mathrm{p} = \omega_0 - \Omega_0$ and a weak probe signal with $\omega_\mathrm{pr} \approx \omega_0$ will interfere inside the SQUID cavity and generate an amplitude beating with the difference frequency $\Omega = \omega_\mathrm{p} - \omega_\mathrm{pr}$.
When $\Omega \approx \pm\Omega_0$, the LF resonator is driven to coherent oscillations by the parametric interaction between the oscillators.
This, in turn, modulates the SQUID cavity resonance frequency and hereby generates a sideband to the pump tone at $\omega = \omega_\mathrm{p} + \Omega$, which interferes with the original probe tone.
The interference effect of PPIT is experimentally identified by a narrow transparency window in the SQUID cavity response. 
In the bottom data of Fig.~\ref{fig:strongcoupling}\textbf{d}, the transparency window is visible as a small peak in the center of the cavity absorption dip and its shape is given by the LF resonator response including dynamical backaction from the red-detuned pump field.
When the multi-photon coupling strength $g=\sqrt{n_c}g_0$ is increased, this transparency window grows in magnitude and width, and in typical optomechanical setups such an enhancement of the multi-photon coupling strength is achieved by increasing the number of intracavity photons by an increased strength of the sideband pump tone \cite{Teufel11a}.
In our device, however, we can control the single-photon coupling rate $g_0$ by changing the flux bias value of the SQUID, similar to flux-mediated microwave optomechanics \cite{Shevchuk17, Rodrigues19}.
We are therefore able to enhance the multi-photon coupling rate while keeping the number of intracavity photons constant, cf. Fig.~\ref{fig:strongcoupling}\textbf{b} and \textbf{c}.
With increasing flux bias and correspondingly increasing single-photon coupling rate $g_0$, the transparency window in the SQUID cavity response grows larger in amplitude and width as can be seen in Fig.~\ref{fig:strongcoupling}\textbf{d} from bottom to top, until for the largest flux bias values two distinct, new eigenmodes define the response.
These new modes of the pumped system correspond to hybridized modes between the LF resonance and intracavity field modulations and both modes are approaching a hybridized linewidth of $(\kappa+\Gamma_0)/2$ and a frequency splitting of $2g$.
This hybridization is apparent as an avoided crossing of the two modes, when the pump frequency is iteratively swept through the red sideband as shown for different flux bias points in \textbf{e}.
From theoretical modelling of the reflection response, shown as black lines in Fig.~\ref{fig:strongcoupling}\textbf{d}, we extract the cooperativity $\mathcal{C} = \frac{4g^2}{\kappa\Gamma_0}$ and the multi-photon coupling rate $g$.
The largest value we show here corresponds to a cooperativity $\mathcal{C} \approx 53$ and to a coupling rate $g/\pi = 0.5\,$MHz.
Additional data on the intracavity photon number dependence for a fixed flux bias point with maximum cooperativity $\mathcal{C} \approx 130$ and $g/\pi \approx 1\,$MHz can be found in SM Sec.~S6.
With a calibration of the setup attenuation, we can determine the intracavity photon number $n_c \approx 70$ for the data presented in Fig.~\ref{fig:strongcoupling} and from there the single-photon coupling rate $g_0$.
The results are plotted in panel \textbf{c} as circles and follow closely the theoretically expected line with maximum values corresponding to $g_0/\kappa \sim 0.1$.
We attribute deviations to a frequency dependent system attenuation and a frequency-dependent conversion from pump power to intracavity photon numbers due to cable resonances in our setup.
As our system provides access to the LF resonator response, we can also detect the normal-mode-splitting directly in the LF reflection, cf. SM Sec.~S6.

\subsection*{Observation of photon-pressure-amplified thermal radio-frequency photons}

\begin{figure}
\centerline{\includegraphics[trim = {0.2cm, 15.0cm, 0.2cm, 0.0cm}, clip=True,scale=0.18]{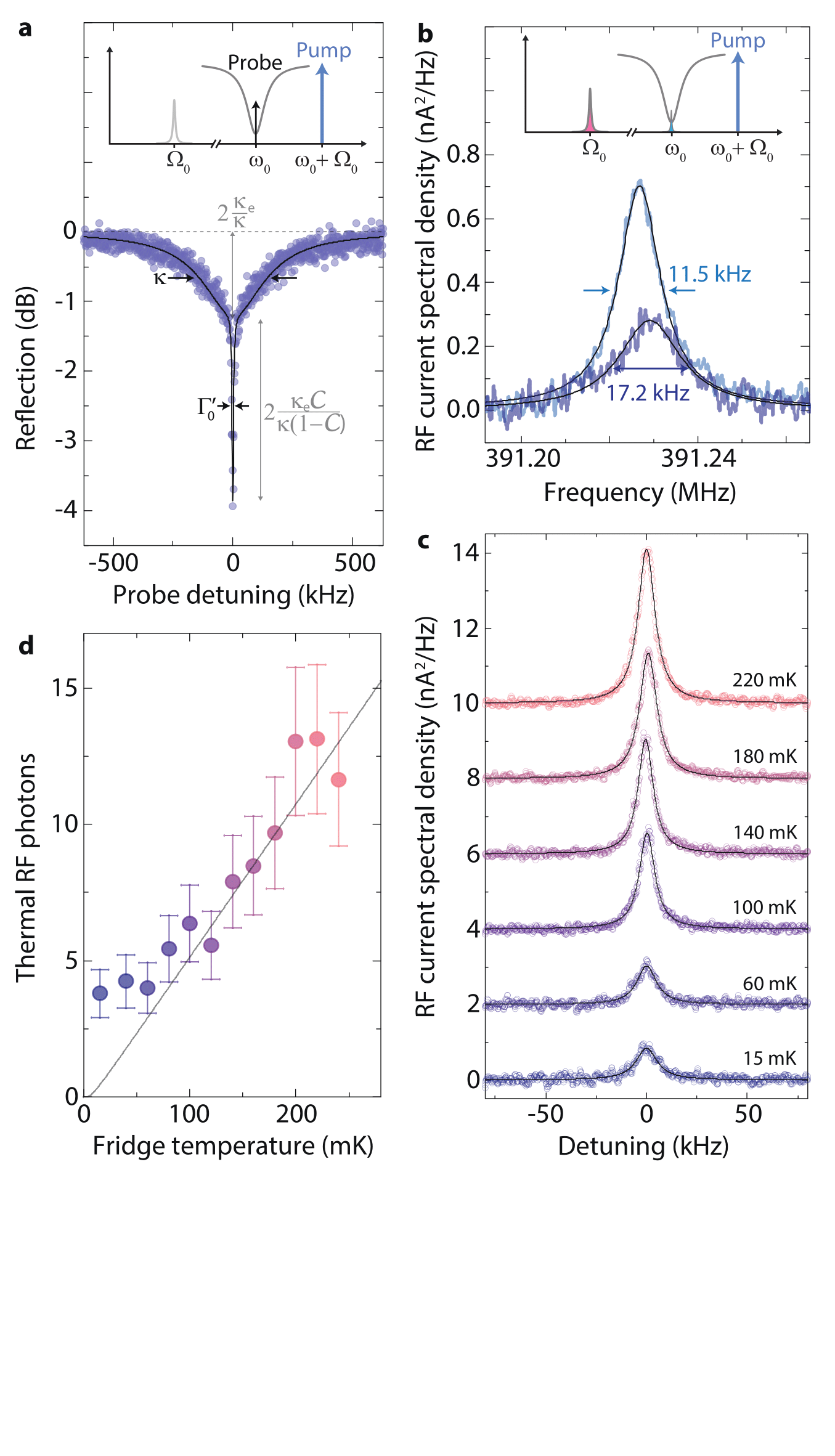}}
\caption{\textsf{\textbf{Observation of photon-pressure-amplified thermal noise of a radio-frequency superconducting circuit.} \textbf{a} When a pump tone is set to the blue cavity sideband $\omega_\mathrm{p} = \omega_0 + \Omega_0$ and a weak probe tone is swept through the cavity resonance (schematic as inset), a narrow absorption window appears in the center of the cavity resonance. From the response, we extract $\kappa, \kappa_e$, the effective LF resonator linewidth $\Gamma_0'$ and $\mathcal{C}$. The flux bias value is $\Phi_b/\Phi_0 \sim 0.5$ and the fridge temperature $T_b=15\,$mK. \textbf{b} Without the probe signal, the residual thermal and quantum fluctuations in the LF resonator are amplified by the blue-detuned tone and generate a sideband to the pump tone at the SQUID cavity center frequency, cf. inset schematic. This noise sideband around the cavity center is detected with a signal analyzer and converted to RF current spectral density. Shwon are two exemplary data sets (black lines are fits) for different photon-pressure amplification gain, the numbers indicate the respective $\Gamma_0'$s. The photon-pressure damping is modified here by slightly detuning the pump from the blue sideband, and a corresponding photon-pressure frequency shift is visible. The LF resonator current noise spectral density for varying fridge temperature is shown as colored lines from $15\,$mK to $220\,$mK together with Lorentzian fits as black lines in \textbf{c}. Subsequent data are manually shifted by $2\,$nA$^2/\,$Hz for better visibility. From the amplitude of the current noise, the equilibrium thermal photon number is calculated and plotted in panel \textbf{d} vs the fridge temperature. Error bars indicate a $20\%$ uncertainty and the gray line follows the Bose-factor. For large fridge temperatures, the RF resonator is thermalised with the fridge, for low fridge temperatures a residual thermal occupation of $\sim 4$ RF photons remains, indicating a mode temperature of $\sim 80\,$mK.}}
\label{fig:Thermal}
\end{figure}
Photon-pressure induced dynamical backaction in parametrically coupled systems does not only influence the resonance frequency and the linewidth of the LF resonator, but at the same time impacts its internal state by cooling or parametric amplification.
This effect has been used to cool mechanical oscillators into the quantum ground-state \cite{Chan11, Teufel11} with a red-detuned pump tone or to realise parametric, mechanical-oscillator-mediated microwave amplification using a blue-detuned pump \cite{Massel11}.
At the base temperature of our dilution refrigerator $T_b = 15\,$mK, the LF resonator is expected to have a thermal photon occupation of $n_\mathrm{th} = \left(e^{\hbar\Omega_0/k_\mathrm{B}T_b}-1\right)^{-1}\approx 0.44$, hence to be close to the quantum ground-state.
A pump tone on the blue SQUID cavity sideband $\omega_\mathrm{p} = \omega_0 + \Omega_0$ will have two effects to the LF resonator.
It will reduce its effective linewidth by negative photon-pressure damping and at the same time it will amplify the intrinsic state of the resonator.
For the observation of the negative damping, we first perform a measurement scheme similar to PPIT in the previous section, but now with a blue-detuned pump tone at $\omega_\mathrm{p} = \omega_0 + \Omega_0$.
The reflection of the SQUID cavity probed with a weak second tone around $\omega_\mathrm{pr} \approx \omega_0$ in presence of a blue-detuned pump is shown in Fig.~\ref{fig:Thermal}\textbf{a}.
Instead of an interference peak as observed for the red-detuned pump field, we now observe a very narrow absorption dip with $\Gamma_0' \approx 2\pi\cdot 10\,$kHz, indicating the regime of photon-pressure induced absorption (PPIA) and negative photon-pressure damping with $\delta\Gamma_0 \approx -2\pi\cdot 12\,$kHz.
From this response curve, we can extract all system parameters such as the linewidths $\kappa, \kappa_e, \Gamma_0'$ and cooperativity $\mathcal{C}$, as indicated in Fig.~\ref{fig:Thermal}\textbf{a}.
To detect the state of the LF resonator, we switch off the probe tone and measure the SQUID cavity output field around its resonance with a signal analyzer in presence of a blue-detuned pump.
The thermal and quantum fluctuations in the LF resonator generate a sideband to the pump, also described as anti-Stokes process in the scattering picture of optomechanics \cite{Aspelmeyer14}, and this noise-induced sideband is detected using a signal analyzer.
For a negligible SQUID cavity occupancy, the detected power spectral density $S(\omega)$, in units of photon number, is related to the current fluctuation spectral density in the LF resonator $S_{I}(\Omega)$ by
\begin{equation}
\frac{S(\omega)}{\hbar\omega} = \frac{1}{2} + n_\mathrm{add}' + \frac{\kappa_e}{\kappa}\frac{\mathcal{C}}{2}\frac{\Gamma_0}{I_\mathrm{zpf}^2}S_I(\Omega)
\end{equation}
with
\begin{equation}
S_I(\Omega) = \frac{8\Gamma_0}{\Gamma_0'^2 + 4\Delta_0^2}(n_\mathrm{LF}+1)
\end{equation}
where $n_\mathrm{LF}+1 = (n_\mathrm{th}+1)/(1-\mathcal{C})$ is the amplified resonator population and $n_\mathrm{add}' \approx 29$ is the effective number of noise photons added by the detection chain.
With these equations, we transform the detected power spectral density into a current fluctuation spectral density, cf. also SM Sec.~S7, and the result is plotted for two different values of photon-pressure amplification in Fig.~\ref{fig:Thermal}\textbf{b}.
We can calibrate the residual thermal occupation of RF photons in the LF resonator by varying the fridge temperature and detect the thermal current spectral density as shown in Fig.~\ref{fig:Thermal}\textbf{c} for six different fridge temperatures.
With increasing temperature, the noise amplitude grows, indicating the increased thermal population of the LF resonator.
For each curve, we determine the effective photon population $n_\mathrm{LF}$ from its amplitude on resonance and estimate the original backaction-free thermal population using $n_\mathrm{th}\approx (1-\mathcal{C}) n_\mathrm{LF} - \mathcal{C}$ with $\mathcal{C}\sim 0.6$.
The resulting thermal occupation is plotted in Fig.~\ref{fig:Thermal}\textbf{c} as circles and shows a trend for higher temperatures, that follows closely the nearly linear behaviour expected from the Bose-factor shown as gray line.
For lower temperatures, the LF mode saturates around $n_\mathrm{th} \approx 4$ thermal photons, which corresponds to a mode temperature of $\sim 80\,$mK. 
In principle, our device also enables sideband-cooling by about a factor of 5 before the strong-coupling regime is reached, therefore enabling cooling the radio-frequency resonator close to its quantum ground-state with $n_\mathrm{LF} < 1$.
Due to the intrinsic sideband asymmetry of the detection scheme, the highly undercoupled SQUID cavity $\kappa_e\kappa \sim 0.1$ and the large number of added photons $n_\mathrm{add}' \approx 29$, however, we are not able to detect the signal within our current setup.

\section*{Discussion}
\vspace{-2mm}

We presented a device consisting of two superconducting circuits, which are coupled via a parametric photon-pressure interaction.
Performing a series of experiments, we demonstrated dynamical backaction between two superconducting circuits, observed photon-pressure-induced transparency and normal-mode splitting, indicating the parametric strong-coupling regime.
Finally, we observed photon-pressure-amplified radio-frequency photons by blue-sideband SQUID cavity pumping.
In summary our device constitutes a novel platform for the control and readout of superconducting quantum circuits of a broad range of frequencies and enables new ways to manipulate and detect radio-frequency photons.
As the system dynamics is completely equivalent to cavity optomechanics, many technological developments of the last decade such as parametric amplifiers, non-reciprocal devices or bath engineered systems, which are based on photon-pressure coupling, can be realised now with a purely circuit-based approach.
Our results also open the door for the investigation of parametrically coupled harmonic oscillators in novel parameter regimes, as superconducting circuits have an extremely high design flexibility and precision regarding resonance frequencies and linewidths compared to opto- or electromechanical systems.
The realization of in particular the strong-coupling regime also enables recently discussed possibilities for quantum computation, using bosonic codes based on GKP states \cite{Gottesmann01, Weigand19}.

\subsection*{Acknowledgements}
\vspace{-2mm}

The authors would like to thank M.~F.~Gely for help with the device fabrication and M.~D.~Jenkins for support with the data acquisition software.
This research was supported by the Netherlands Organisation for Scientific Research (NWO) in the Innovational Research Incentives Scheme -- VIDI, project 680-47-526.
This project has received funding from the European Research Council (ERC) under the European Union's Horizon 2020 research and innovation programme (grant agreement No 681476 - QOMD) and from the European Union's Horizon 2020 research and innovation programme under grant agreement No 732894 - HOT.

\subsection*{Author contributions}
\vspace{-2mm}

DB and ICR designed and fabricated the device, performed the measurements and analysed the data.
GAS conceived the experiment and supervised the project.
DB and ICR edited the manuscript with input from GAS.
All authors discussed the results and the manuscript.

\subsection*{Competing interest}
\vspace{-2mm}
The authors declare no competing interests.

\newpage

\widetext

\noindent\textbf{\textsf{\Large Supplementary Material: Photon-Pressure Strong-Coupling between \\ two Superconducting Circuits}}

\normalsize
\vspace{.3cm}

\noindent\textsf{D.~Bothner$^*$, I.~C.~Rodrigues$^*$ and G.~A.~Steele}

\vspace{.2cm}
\noindent\textit{Kavli Institute of Nanoscience, Delft University of Technology, PO Box 5046, 2600 GA Delft, The Netherlands\\$^*$\normalfont{these authors contributed equally}}

\renewcommand{\thefigure}{S\arabic{figure}}
\renewcommand{\theequation}{S\arabic{equation}}

\renewcommand{\thesection}{S\arabic{section}}
\renewcommand{\bibnumfmt}[1]{[S#1]}
\renewcommand{\citenumfont}[1]{S#1}

\setcounter{figure}{0}
\setcounter{equation}{0}

\section{Device fabrication}
\label{Section:Fab}

\begin{figure}[h]
	\centerline {\includegraphics[trim={0.5cm 5.5cm 0.5cm 0cm},clip=True,scale=0.75]{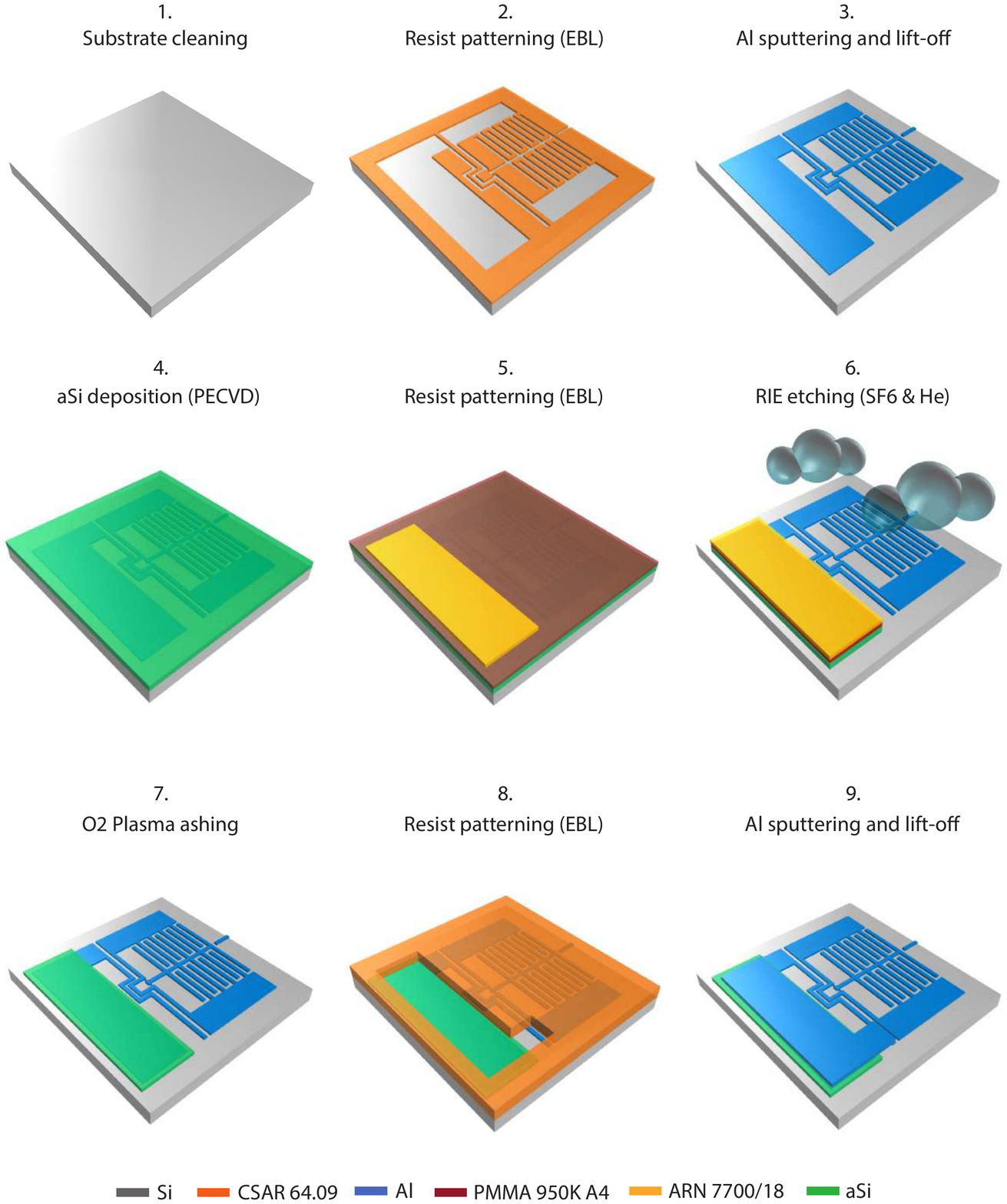}}
	\caption{\textsf{\textbf{Schematic device fabrication.} 1.-3. Show the deposition and patterning of the SQUID cavity and both bottom plate and inductor wire of the low frequency cavity 4.-6. Deposition and patterning of the dielectric layer for the parallel plate capacitor 7.-8. Patterning and deposition of the top plate of the low-frequency circuit capacitor. 9. Final device. A detailed description of the individual steps is given in the text.}}
	\label{fig:Fab}
\end{figure}
The device fabrication starts with the patterning of alignment markers on a full $4\,$inch Silicon wafer, required for the EBL (Electron Beam Lithography) alignment of the following fabrication steps.
The structures were patterned using a CSAR62.13 resist mask and a sputter deposition of $50\,$nm Molybdenum-Rhenium alloy. After undergoing a lift-off process, the only remaining structures on the wafer were the markers.
The complete wafer was diced into individual $14\times14\,$mm$^2$ chips, which were used individually for the subsequent fabrication steps.
As second step in the fabrication, we pattern the bottom plate of the parallel plate capacitor, the inductor wire of the low-frequency cavity, the SQUID cavity (Fig.~\ref{fig:Fab} 2.$\ \&$ 3.) and the center conductor of the SQUID cavity feedline by means of EBL using CSAR62.09 as resist. 
After the exposure, the sample was developed in Pentylacetate for $60\,$seconds, followed by a solution of MIBK:IPA (1:1) for $60\,$seconds, and finally rinsed in IPA. 
The sample was subsequently loaded into a sputtering machine where a $20\,$nm layer of Aluminum was deposited.
Finally, the chip was placed in the bottom of a beaker containing a small amount of anisole and inserted in a ultrasonic bath for a few minutes.
This lift-off process turned out to be very efficient compared to warm anisole bath without ultrasound, where sometimes some of the unwritten structures would not lift-off. 
The deposition of the dielectric layer of the parallel plate capacitors was done using a plasma-enhanced chemical vapor deposition (PECVD).
To guarantee low dielectric losses in the material, the chamber underwent an RF cleaning process overnight and only afterwards the deposition of $\sim130\,$nm of amorphous silicon was performed.
At this point the whole sample is covered with dielectric, cf. Fig.~\ref{fig:Fab} 4.
Afterwards, a double layer of resist (PMMA 950K A4 and ARN-7700-18) was spin-coated and exposed with EBL. 
Prior to the development of the pattern, a post-bake of 2 minutes at $\sim115\,^{\circ}{\rm C}$ was required.
Directly after, the sample was dipped in MF321 for 2 minutes and 30 seconds, followed by H$_2$O for 30 seconds and lastly rinsed in IPA, cf. Fig.~\ref{fig:Fab} 5.
To finish the third step of the fabrication, the developed sample underwent a SF$_6$/He reactive ion etching (RIE) to remove the amorphous Silicon, followed by a O$_2$ plasma ashing to remove resist residues, cf. Fig.~\ref{fig:Fab} 6.$\ \&$ 7.
As final step, the sample was again coated in CSAR62.13 and the top plate of the capacitor as well as all ground plane and the low-frequency feedline was patterned with EBL.
The development of the resist was done in a similar way to the one mentioned in the second step.
Afterwards, the sample was loaded into a sputtering machine where an argon milling process was performed in situ, for 2 minutes.
This etching step prior to the deposition was done to remove the native aluminum oxide present on top of the first layer and allow for good electrical contact between the top and bottom plates of the low-frequency capacitor.
Posterior to the milling, a $200\,$nm layer of Aluminum was deposited and finally a lift-off procedure, similar to the one of the second step, was performed, cf. Fig.~\ref{fig:Fab} 9.
At the end of the fabrication, the sample was diced to a $10\times10\,$mm$^2$ size and mounted into a PCB (Printed Circuit Board).
A schematic representation of this fabrication process can be seen in Fig.~\ref{fig:Fab}, omitting the initial patterning of the electron beam markers.
\begin{figure}
	\centerline {\includegraphics[trim={0cm 0cm 0cm 0.5cm},clip=True,scale=0.8]{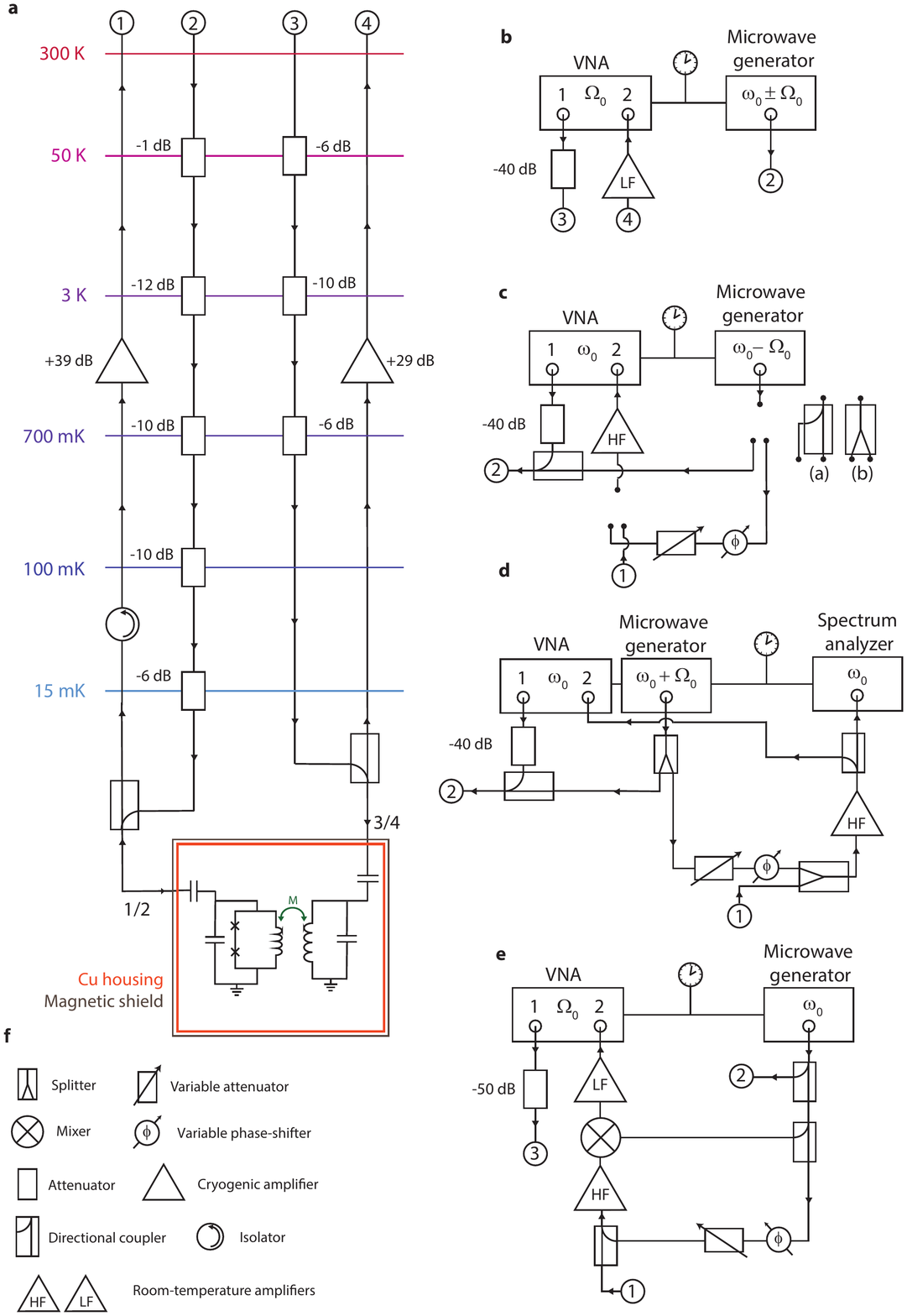}}
	\caption{\textsf{\textbf{Schematic of the measurement setup.} Detailed information is provided in text.}}
	\label{fig:Setup}
\end{figure}

\section{Measurement setup}

\subsection{Setup configuration}

All the experiments reported in this paper were performed in a dilution refrigerator operating at a base temperature close to $T_b = 15\,$mK.
A schematic of the experimental setup and of the external configurations used in the different performed experiments can be seen in Fig.~\ref{fig:Setup}.

The PCB (Printed Circuit Board), onto which the fabricated sample was glued and wirebonded, was placed in a radiation tight copper housing and connected to two coaxial lines. 
One of the lines was used as input/output port for the high-frequency SQUID cavity and the second line was set in a similar way for the low-frequency cavity.
Both of the cavities were measured in a reflection geometry, and therefore the input and output signals were split via a directional coupler on the $15\,$mK stage. 
Both output signals went into a cryogenic HEMT amplifier for their particular frequency range.
Furthermore, in order to generate an out-of-plane magnetic field, required to flux bias the SQUID cavity, an external magnet (not shown in the figure) was put in very close proximity below the device and the two were placed inside a magnetic shield.
The magnet was connected with DC wires, allowing for the field to be tuned by means of a DC current (not shown).
Both high-frequency input lines were heavily attenuated in order to balance the thermal radiation from the line to the base temperature of the fridge. 
Outside of the refrigerator, we used different configurations of microwave signal sources and high-frequency electronics for the different experiments.
In \textbf{b} we show the configuration of the measurement of dynamical backaction.
A microwave generator sends a continuous wave signal to the SQUID cavity around one of its sidebands, while the LF resonator is probed in reflection with a vector network analyzer (VNA). 
We also used this setup to measure the normal-mode splitting on the LF side as shown in Fig.~\ref{fig:LFSC}.
In \textbf{c} the setup for photon-pressure induced transparency and the strong-coupling regime detected at the SQUID cavity is shown. 
During the experiment a continuous wave tone from a microwave generator is combined with a weak probe signal via a directional coupler and sent to the SQUID cavity.
The output signal, coming from the dilution refrigerator, is afterwards carrier-cancelled with the original pump tone in order to avoid reaching the saturation regime of our room temperature (RT) amplifiers, and analyzed by means of a VNA.
For the data shown in the main paper, a directional coupler was used at the output of the microwave generator and before the RT high-frequency amplifier.
In order to drive the SQUID cavity with higher powers for the data shown in Fig.~\ref{fig:PowerStrongCoupling}, the directional couplers, which have an intrinsic loss of $10\,$dB, were replaced by splitters.
The two configurations are denoted (a) for the directional coupler case and (b) for the splitter/combiner case.
In \textbf{d} we show the setup for photon-pressure induced absorption and thermal noise amplification and detection (main paper Figs.~4\textbf{a}, \textbf{c} and \textbf{d}), where a continuous tone is send to the blue sideband of the SQUID cavity.
In addition, in order to observe the cavity response and adjust the pump tone frequency with respect to the power-dependent cavity resonance, a weak VNA signal is combined with the pump tone via a directional coupler.
The output signal is then carrier-cancelled with the original pump tone, amplified and split in two signals that are analyzed individually by a spectrum analyzer and a VNA.
During the detection of thermal noise with the signal analyzer, the VNA scan was stopped and the VNA output power was completely switched off.
In \textbf{e}, and the setup for LF resonance up-conversion is shown (main paper data in Fig.~1\textbf{e}).
During the experiment, a resonant pump tone is sent to the SQUID cavity using a signal generator.
Simultaneously a weak scanning probe signal coming from the VNA is sent to the LF resonator.
The SQUID cavity output field is carrier-cancelled with the original pump tone, amplified, down-converted and once again amplified with a LF amplifier before reaching the VNA input port.
For the data shown in the main paper Fig.~4\textbf{b}, we used a combination of setups \textbf{d} and \textbf{e}.
Everything is identical to \textbf{d}, but instead of directly detecting the high-frequency sideband signal after the final directional coupler, we added a mixer with the pump tone as local oscillator for down-conversion.
Also, we added another LF amplifier before the signal was entering the signal analyzer.
This way, we detected the LF thermal noise directly at its original oscillation frequency.
For all experiments, the microwave sources and vector network analyzers (VNA) as well as the spectrum analyzer used a single reference clock of one of the devices.

\subsection{Estimation of the attenuation chain}

In order to estimate the input power on the on-chip feedline of the device, we use the thermal noise of the HEMT amplifier as calibration method.
The cryogenic HEMT (High Electron Mobility Transistor) amplifier thermal noise power is given by
\begin{equation}
P_\mathrm{HEMT} = 10\,\log\left(\frac{k_\mathrm{B} T_{\mathrm{HEMT}}}{1\,\textrm{mW}}\right) + 10\,\log\left(\frac{\Delta f}{\textrm{Hz}}\right)
\label{eq:HEMT}
\end{equation}
where $k_\mathrm{B}$ is the Boltzmann constant, $T_{\mathrm{HEMT}}$ is the noise temperature of the amplifier, which, according to the specification datasheet, is approximately 5.5 K, and $\Delta f = 200\,$Hz is the measurement IF bandwidth.
The calculated noise power is $P_\mathrm{HEMT} = -168.2\,$dBm, or as noise RMS voltage $\Delta V = 0.87\,$nV.

For a VNA output power of $-30\,$dBm, we extract a signal-to-noise ratio $\textrm{SNR}=46.1\,$dB.
From this we extract a total power arriving at the HEMT of $-122\,$dBm.
Furthermore, assuming an attenuation between the sample and the HEMT of $2\,$dB, and taking into account the room temperature attenuators (29 dB) used at the output of the VNA, we estimate a total input attenuation in the line of $-61\,$dB.
This calibration method was repeated for different frequency ranges, allowing us to extract an error of $\pm 2\,$dB. 
Since the probe and the pump were always combined outside the dilution fridge and sent via the same input line, during the experiment an attenuation $G =-61\,$dB was used for both tones.

\section{The LF resonator}

\subsection{Analytical circuit model}
The LF resonator used in this experiment is a simple LC circuit coupled to a feedline with characteristic impedance $Z_0$.
The resonator consists of a parallel plate capacitor with an area of $A_\mathrm{LF} = 7.68\cdot10^{-7}\,$m$^2$, filled with $t \approx 130\,$nm thick amorphous silicon as dielectric.
With
\begin{equation}
C_\mathrm{LF} = \epsilon_0\epsilon_r\frac{A_\mathrm{LF}}{t}
\end{equation}
the capacitance is calculated to be $C_\mathrm{LF} \approx 620\,$pF, where $\epsilon_0 = 8.854\cdot10^{-12}\,$F/m is the vacuum permittivity and $\epsilon_r = 11.8$ is the relative permittivity of silicon.
In addition, the resonator is capacitively coupled to a coplanar waveguide feedline by means of a parallel plate coupling capacitor with $C_c = 434\,$fF.
From the resonance frequency of $\Omega_0 = 2\pi\cdot391\,$MHz and using
\begin{equation}
\Omega_0 = \frac{1}{\sqrt{L_\mathrm{LF}(C_\mathrm{LF} + C_c)}}
\end{equation}
we calculate the total inductance of the circuit as $L_\mathrm{LF} = 267\,$pH.
This effective inductance includes the contribution from the mutual inductive coupling of the two circuits.
A significant contribution to the total inductance is coming from kinetic inductance due to the thin film used for the bottom layer and the inductor wire.
From corresponding simulation with the software package SONNET, we estimate the kinetic inductance $L_k \sim 2.2\,$pH/sq to contribute about $0.59$ to the total inductance or $L_k = 1.44L_g$ with the kinetic inductance $L_k$ and the geometric inductance $L_g$.
For the capacitively coupled parallel LC circuit, the external linewidth is given by
\begin{equation}
\Gamma_e = \frac{Z_0 C_c^2}{L_\mathrm{LF}(C_\mathrm{LF}+C_c)^2}
\end{equation}
which gives with $Z_0 = 50\,\Omega$ a value of $\Gamma_e = 2\pi\cdot 14.5\,$kHz.

\subsection{Current zero-point fluctuations}
The current zero-point fluctuations through the LF inductor are calculated via
\begin{equation}
I_\mathrm{zpf} = \sqrt{\frac{\hbar\Omega_0}{2L_\mathrm{LF}}}
\end{equation}
and with the parameters of the circuit we get $I_\mathrm{zpf} \approx 21\,$nA.

\subsection{Cavity parameter extraction from a reflection measurement}
\begin{figure}
	\centerline {\includegraphics[trim={0cm 21.5cm 0cm 0cm},clip=True,width=0.8\textwidth]{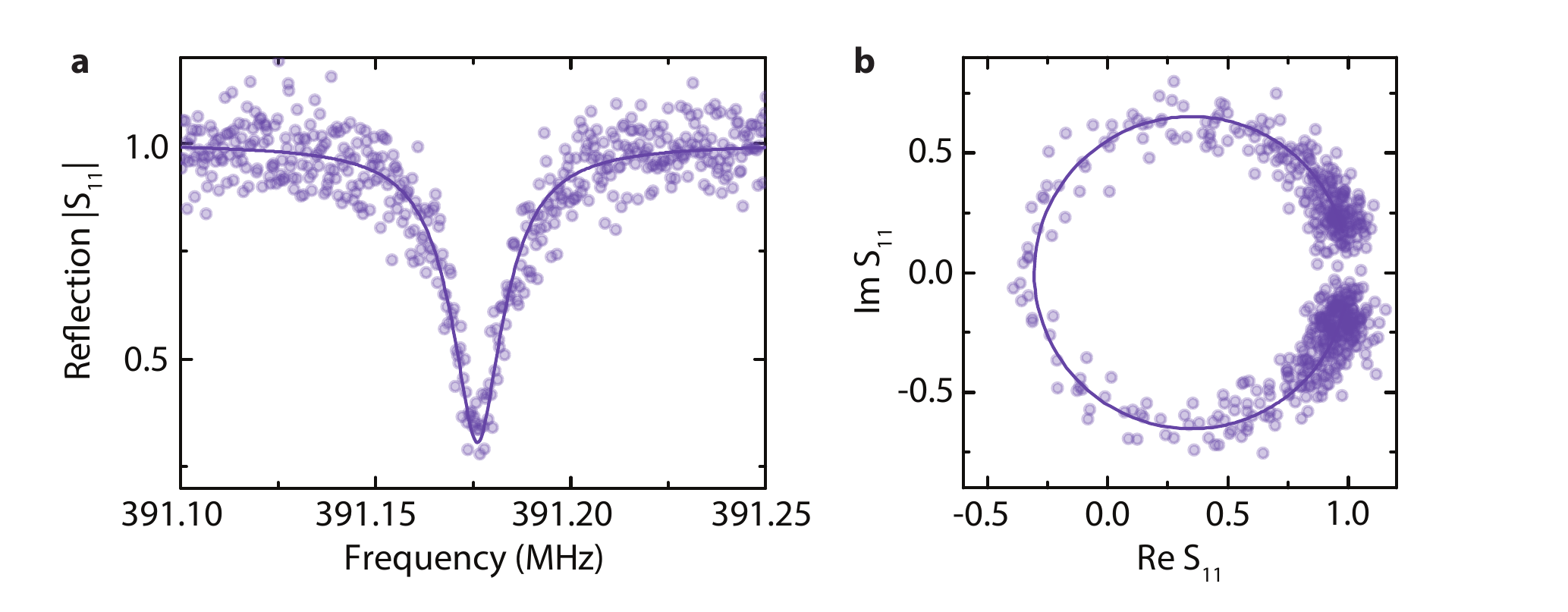}}
	\caption{\textsf{\textbf{The LF resonator response.} \textbf{a} The magnitude of the reflection response $|S_{11}|$ of the low-frequency resonator. Circles are data, line is a fit. \textbf{b} The complex scattering data $S_{11}$ around the LF resonance frequency. Circles are data, line is a fit. The data and fits shown here have been background-corrected before plotting as described in Sec.~S5. From the fit, we extract $\Gamma_e, \Gamma_i, \Omega_0$ as given in the text.}}
	\label{fig:LFfit}
\end{figure}
Figure~\ref{fig:LFfit} shows a background corrected LF resonator reflection measurement in amplitude data $|S_{11}|$ in \textbf{a} as well as in the complex plane in \textbf{b}, together with the corresponding fit curves.
From fitting the data with the response function
\begin{equation}
S_{11} = 1 - \frac{2\Gamma_e}{\Gamma_i + \Gamma_e +2i\Delta_0}
\end{equation}
with $\Delta_0 = \Omega - \Omega_0$ and the internal and external decay rates $\Gamma_i$ and $\Gamma_e$, we extract the resonator parameters
\begin{equation}
\Omega_0 = 2\pi\cdot 391.18\,\mathrm{MHz}, ~~~~~ \Gamma_i = 2\pi\cdot 7.4\,\mathrm{kHz}, ~~~~~ \Gamma_e = 2\pi\cdot 13.8\,\mathrm{kHz}
\end{equation}
where the extracted $\Gamma_e$ is very close to the theoretical value of $14.5\,$kHz.

\section{The HF SQUID cavity}

\subsection{Analytical circuit model}

\subsubsection*{Capacitance, inductance, and feedline coupling}

The capacitance of an interdigitated capacitor (IDC) can be approximately calculated as given in Ref.~\cite{Igreja04}, where
\begin{equation}
C_\mathrm{IDC} = (N-3)\frac{C_1}{2} + 2\frac{C_1C_2}{C_1 + C_2}
\end{equation}
with
\begin{equation}
C_i = 2\epsilon_0\epsilon_\mathrm{eff}l\frac{K(k_i)}{K(k_i')}, ~~~~~ i = 1,2.
\end{equation}
Here, $K(k_i)$ are elliptic integrals of the first kind, $l$ is the finger length, $\epsilon_\mathrm{eff} = (\epsilon_r+1)/2$ is the effective permittivity with the Silicon substrate permittivity $\epsilon_r =11.8$, $N$ is the total number of fingers and
\begin{eqnarray}
k_1 & = & \sin\left(\frac{\pi}{2}\frac{a}{a+b}\right)\\
k_2 & = & 2\frac{\sqrt{a(a+b)}}{2a + b}\\
k_i' & = & \sqrt{1-k_i^2}
\end{eqnarray}
with $a$ the finger width and $b$ the gap width in between two fingers.
For a single capacitor $C_\mathrm{IDC}$ of our circuit with $N = 90$, $l = 100\,\mu$m, $a = b = 1\,\mu$m, we get $C_\mathrm{IDC} = 507\,$fF.
Since we have two of these capacitors in parallel, the total circuit capacitance is $C_\mathrm{HF} = 2C_\mathrm{IDC} = 1.01\,$pF.
The ground side of each of the two IDCs is not galvanically connected to the ground plane but via a large parallel plate capacitor with $C_\mathrm{pp}\sim 70\,$pF, in order to avoid the generation of a closed superconducting loop around the SQUID, which would act as flux transformer and induce flux-induced frequency noise to the cavity.
Furthermore, the coupling capacitor to the feedline is provided by a $1\,\mu$m gap between the feedline and the cavity and it is estimated to be $C_c'\sim 2\,$fF.
The resonance frequency of the circuit is $\omega_0 = 2\pi\cdot 5.844\,$GHz and related to the circuit parameters by
\begin{equation}
\omega_0 = \frac{1}{\sqrt{L_\mathrm{HF}({C_\mathrm{HF}+C_c'})}}
\end{equation} 
which gives a total inductance of $L_\mathrm{HF} = 742\,$pH at the cavity sweet spot.
This effective inductance is composed of the linear inductors $L_0$, the loop inductance $L_{l}$, the junction inductances $L_\mathrm{J}$ (cf. main paper Fig.~1\textbf{a}) and includes the contribution from the mutual inductive coupling of the two circuits.
The external linewidth is given by
\begin{equation}
\kappa_e = \frac{Z_0 C_c'^2}{L_\mathrm{HF}(C_\mathrm{HF}+C_c')^2}
\end{equation}
and with $C_c' = 2\,$fF results in $\kappa_e = 2\pi\cdot43\,$kHz.

\subsubsection*{SONNET simulations and kinetic inductance}
The total inductance of both cavities has a significant contribution from the kinetic inductance of the $20\,$nm thick Aluminum film.
The kinetic inductance contribution is estimated from simulations using the software package SONNET, where we match the bare cavity resonance frequency with the experimental value by tuning the kinetic inductance per square $L_\Box$.
For a vanishing surface impedance we find a resonance frequency $\omega_{0} = 2\pi\cdot 10.3 \,$GHz in the simulations and achieve high agreement with the experimental value of $\omega_0 = 2\pi\cdot 5.844\,$GHz when $L_\Box =  2.2\,$pH/sq.
Using this kinetic inductance for the bottom plate and the inductor wire of the LF resonator, the LF resonance frequency is simultaneously shifted from $\Omega_0' = 2\pi\cdot 610 \,$MHz for zero kinetic inductance to $\Omega_0 \approx 2\pi\cdot 390\,$MHz.

\subsection{Cavity parameter extraction from a reflection measurement}
\begin{figure}
	\centerline {\includegraphics[trim={0cm 21.5cm 0cm 0cm},clip=True,width=0.8\textwidth]{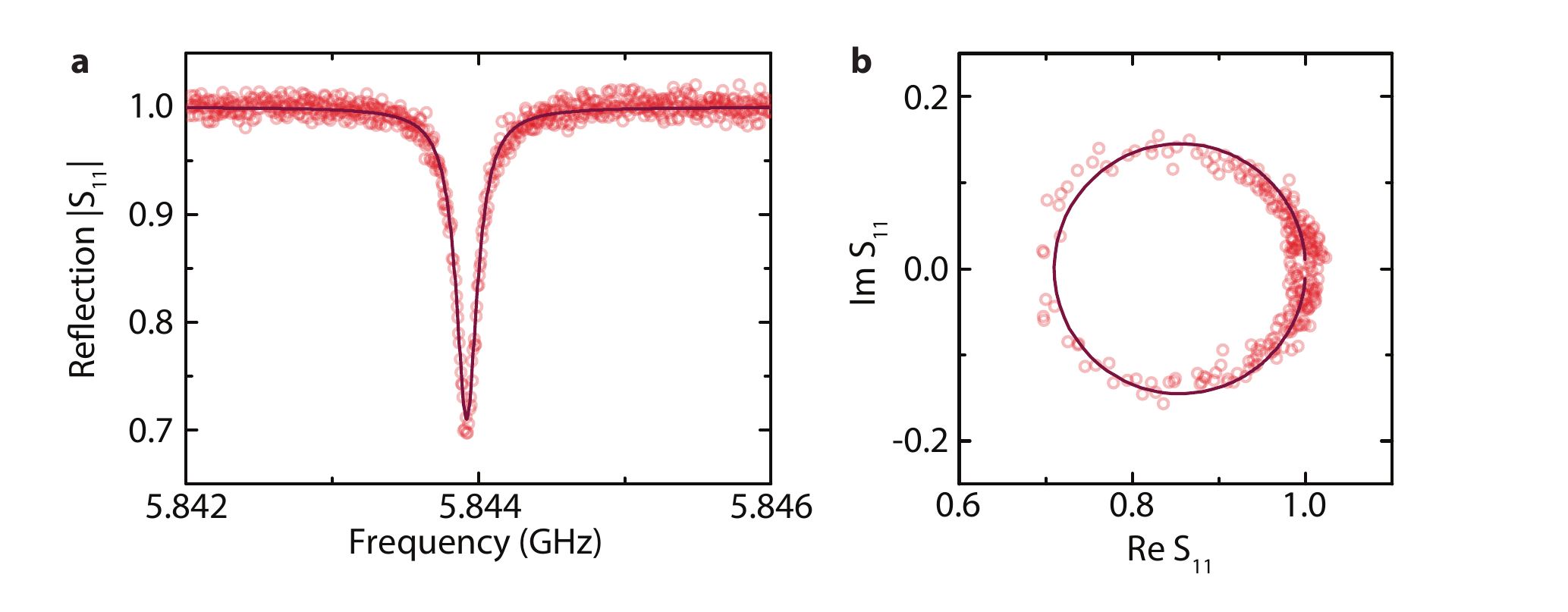}}
	\caption{\textsf{\textbf{The SQUID cavity response.} \textbf{a} The magnitude of the reflection response $|S_{11}|$ of the high-frequency SQUID cavity. Circles are data, line is a fit. \textbf{b} The complex scattering data $S_{11}$ around the HF resonance frequency. Circles are data, line is a fit. The data and fits shown here have been background-corrected before plotting as described in Sec.~S5. From the fit, we extract $\kappa_e, \kappa_i, \omega_0$ as given in the text.}}
	\label{fig:HFfit}
\end{figure}
Figure~\ref{fig:HFfit} shows a SQUID cavity reflection measurement in amplitude data $|S_{11}|$ in \textbf{a} as well as in the complex plane in \textbf{b} and the corresponding fit curves as lines.
From fitting the data with the response function
\begin{equation}
S_{11} = 1 - \frac{2\kappa_e}{\kappa_i + \kappa_e +2i\Delta}
\end{equation}
where $\Delta = \omega - \omega_0$ and the internal and external decay rates are $\kappa_i$ and $\kappa_e$, we extract the parameters
\begin{equation}
\omega_0 = 2\pi\cdot 5.844\,\mathrm{GHz}, ~~~~~ \kappa_i = 2\pi\cdot 163\,\mathrm{kHz}, ~~~~~ \kappa_e = 2\pi\cdot 28\,\mathrm{kHz}
\end{equation}
The extracted $\kappa_e$ slightly deviates from the theoretical value of $43\,$kHz, probably due to a coupling capacitance of $C_c' \approx 1.5\,$fF, which is lower than expected.
The deviation of $\kappa_i + \kappa_e$ from the value given in the main paper is explained by a different power used in both experiments.
We observe that the internal linewidth is flux- and power-dependent as detailed in Sec.~S4.

\subsubsection*{The constriction type Josephson junctions, the SQUID and $\beta_L$}
The constriction type Josephson junctions in our cavity are designed to be $150\,$nm long and $50\,$nm wide and we estimate their critical current to be $I_c \sim 10\,\mu$A, which relates to a Josephson inductance of a single junction of $L_\mathrm{J0} = \Phi_0/2\pi I_c = 32\,$pH.
The SQUID loop is a $10\times10\,\mu$m$^2$ large loop made from a $1\,\mu$m wide wire.
From the total inductance per square $L_\Box \approx 3\,$pH/sq, we estimate the loop inductance to be $L_l = 120\,$pH.
From these numbers, we find the screening parameter of the SQUID as approximately $\beta_L = 2L_lI_c/\Phi_0 = 1.2$.

\subsection{SQUID cavity flux dependence}
\subsubsection*{Resonance frequency}
\begin{figure}
	\centerline {\includegraphics[trim={0cm 21cm 0cm 0cm},clip=True,width=0.98\textwidth]{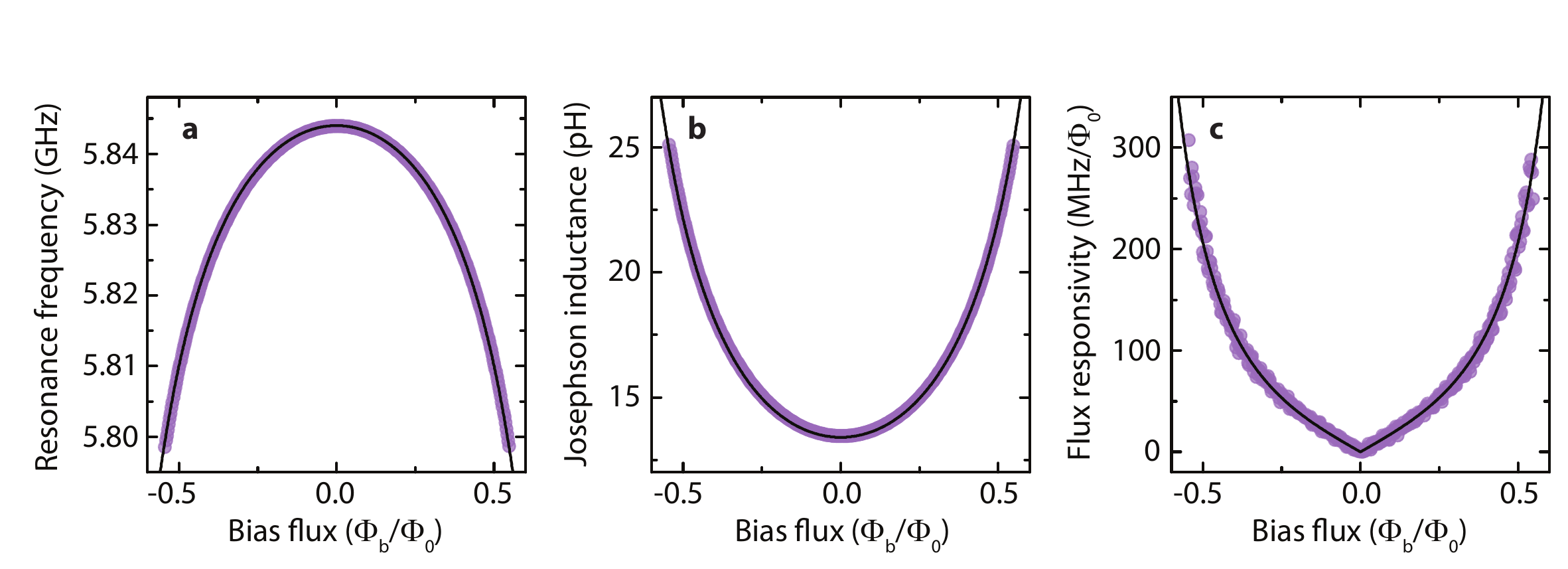}}
	\caption{\textsf{\textbf{Tuning the SQUID cavity with magnetic flux.} \textbf{a} SQUID cavity resonance frequency vs magnetic flux. Circles are data, the black line is a fit. \textbf{b} From the fit of the resonance frequency and with the total inductance $L_\mathrm{HF} = 742\,$pH, the Josephson inductance of the SQUID can be extracted. Circles are extracted from data, the line is a calculation based on the fit parameters from \textbf{a}. The change of resonance frequency with magnetic flux $\partial\omega_0/\partial\Phi$ is shown in \textbf{c}. Both, data (circles) and fit (line) are obtained by differentiating the corresponding curves plotted in \textbf{a}.}}
	\label{fig:FluxArch}
\end{figure}

To take into account the non-negligible SQUID loop inductance and a possible non-sinusoidal current-phase relation, both leading to flux multistability and a widening of the flux arch \cite{Vijay10, Kennedy19}, we phenomenologically describe the SQUID critical current dependence on magnetic flux as
\begin{equation}
I_c(\Phi_b) = 2I_c\cos{\left(\pi\gamma_L\frac{\Phi_b}{\Phi_0}\right)}
\end{equation}
where $\gamma_L$ is a parameter taking into account the widening of the flux arch.
This relates to the Josephson inductance of the SQUID as
\begin{equation}
L_\mathrm{JJ} = \frac{L_\mathrm{J0}}{2\cos{\left(\pi\gamma_L\frac{\Phi_b}{\Phi_0}\right)}}.
\end{equation}
The factor of 2 in the denominator originates from the two junctions in parallel in the SQUID.
With this, the resonance frequency of the SQUID cavity is approximately given by
\begin{equation}
\omega_0(\Phi_b) = \frac{\omega_0(0)}{\sqrt{\Lambda + \frac{1-\Lambda}{\cos{\left(\pi\gamma_L\frac{\Phi_b}{\Phi_0}\right)}}}}
\label{eqn:FluxDep}
\end{equation}
where $\Lambda = (L_\mathrm{HF}-\frac{1}{2}L_\mathrm{J0})/L_\mathrm{HF}$.
Figure \ref{fig:FluxArch}\textbf{a} shows the experimentally determined flux dependence of the resonance frequency together with a fit line using Eq.~(\ref{eqn:FluxDep}).
From this fit, we extract the parameters $\Lambda = 0.982$ and $\gamma_L = 0.59$, which corresponds to a single junction Josephson inductance of approximately $L_\mathrm{J0} = 27\,$pH, i.e., to a critical current of a single junction of $I_c = 12\,\mu$A.
The SQUID Josephson inductance dependent on magnetic flux as extracted from the cavity fit is plotted in panel \textbf{b}.
In panel \textbf{c}, we plot the flux responsivity of the cavity $|\partial\omega_0/\partial\Phi|$, obtained as derivative from \textbf{a}, which is directly related to the coupling rate $g_0$ between the circuits as discussed below.
The maximum responsivity we obtain here, is $|\partial\omega_0/\partial\Phi| \approx 2\pi\cdot 300\,$MHz/$\Phi_0$.

\subsubsection*{Flux axis calibration}

\begin{figure}
	\centerline {\includegraphics[trim={4cm 20.2cm 3cm 0cm},clip=True,width=0.65\textwidth]{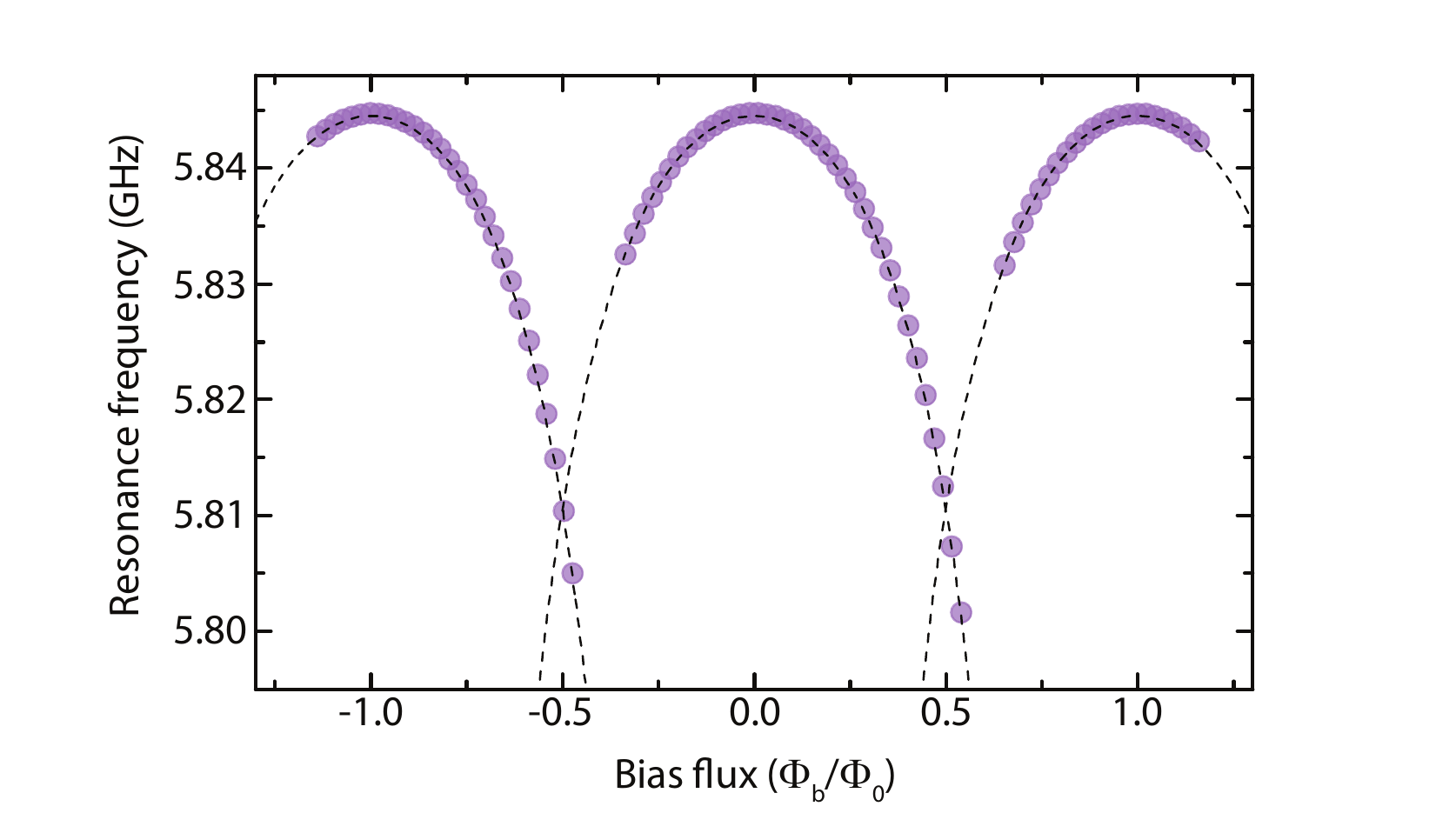}}
	\caption{\textsf{\textbf{Tuning the SQUID cavity with magnetic flux over multiple flux quanta.} Circles show the SQUID cavity resonance frequencies extracted from fits to the measurement data. Some points for very low frequencies $\omega_0 < 2\pi\cdot 5.8\,$GHz are not included, because the fit failed due to extremely shallow resonance dips at high bias flux values. The dashed lines are copies of the fit curve shown in Fig.~\ref{fig:FluxArch}\textbf{a}, one is shifted by $+\Phi_0$, one by $-\Phi_0$ on the flux axis.}}
	\label{fig:FluxArches}
\end{figure}

The flux axis is calibrated by measuring the SQUID cavity resonance frequency over a larger flux range and using a periodicity of one flux quantum.
Figure~\ref{fig:FluxArches} shows the SQUID cavity resonance frequency for a larger flux range with three different archs.
The dashed line are copies of the fit function from Fig.~\ref{fig:FluxArch}\textbf{a}, two of them have been shifted by $-\Phi_0$ and $+\Phi_0$, respectively.

\subsection{Power dependence and anharmonicity of the SQUID cavity}
\label{sec:PowDep}

\subsubsection*{Power dependence of cavity parameters}

We observe that the cavity resonance frequency, due to the Kerr nonlinearity, as well as the internal cavity linewidth, due to nonlinear dissipation, depend on the intracavity photon number $n_c$.
Both, as well as very slightly the external linewidth, depend, in addition, on the flux bias.
\begin{figure}[h]
	\centerline {\includegraphics[trim={0cm 17cm 0cm 1.5cm},clip=True,scale=0.75]{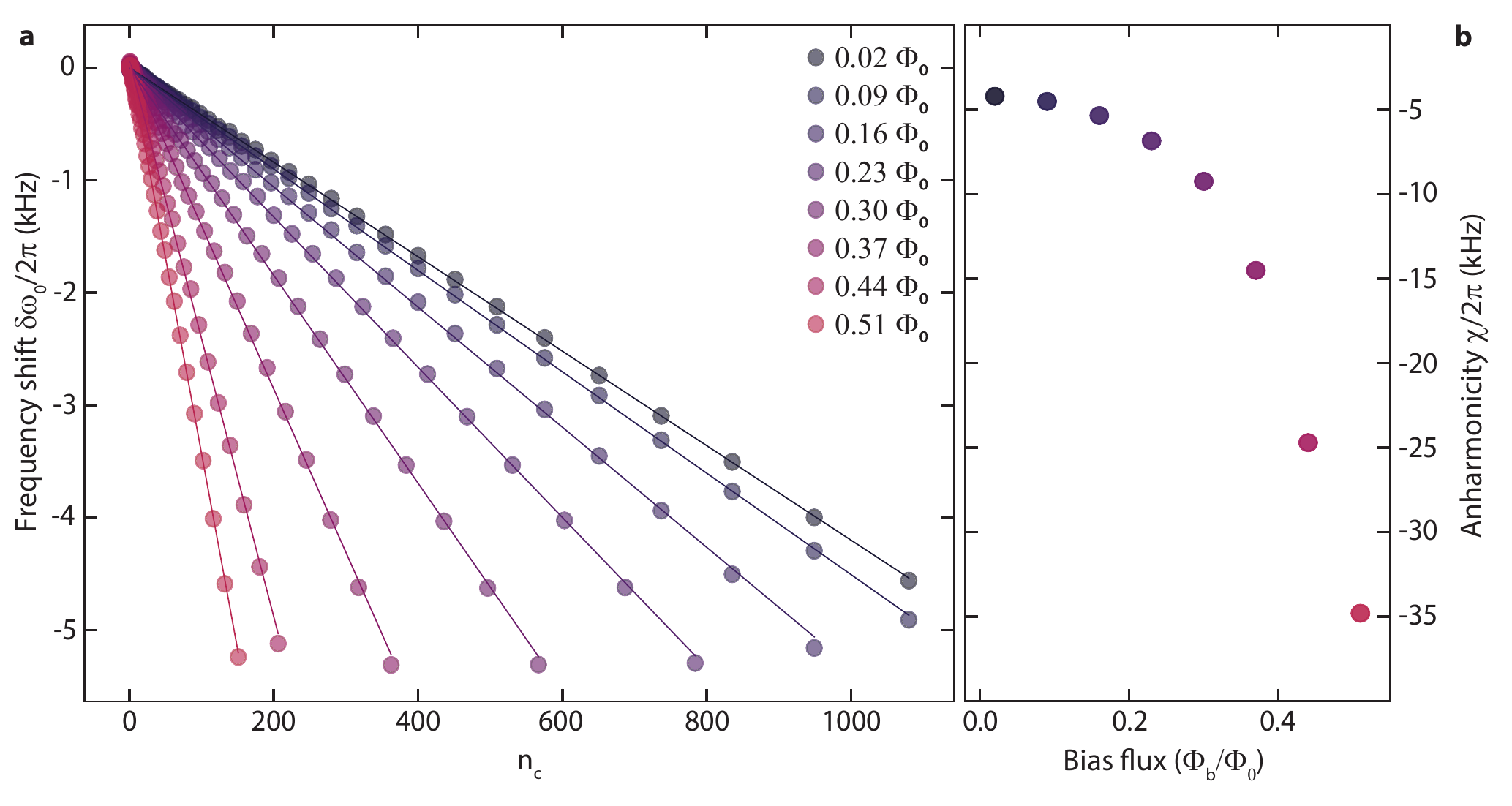}}
	\caption{\textbf{Characterizing the SQUID cavity Kerr nonlinearity.} \textbf{a} Resonance frequency shift $\delta\omega_0 = \omega_0(n_c) - \omega_0(0)$ depending on intracavity photon number, for different flux bias points. Circles are data, lines are linear fits. From the linear fits, we extract the frequency shift per photon for each bias flux value, as plotted as circles in \textbf{b}.}
	\label{fig:PowDepFreq}
\end{figure}
To measure the photon number dependence of $\omega_0$ and $\kappa$, we bias the cavity at a desired flux value and send a microwave pump tone to the cavity around $\omega_p(\Phi_b) = \omega_0(\Phi_b) - \Delta_p$ with $\Delta_p \approx 2\pi\cdot 130\,$MHz, chosen at a detuning which avoids interaction with the LF resonator.
Then, we measure the cavity response with a weak probe tone and fit the curves to extract $\omega_0$, $\kappa_i$ and $\kappa_e$.
Figure~\ref{fig:PowDepFreq} \textbf{a} shows the pump-induced frequency shift $\delta\omega_0$ for different flux bias values vs the calculated intracavity photon number.
The photon number is calculated using
\begin{equation}
n_c = \frac{4P_\mathrm{in}}{\hbar\omega_\mathrm{p}}\frac{\kappa_e}{\kappa^2 + 4\Delta_1^2}
\label{eqn:nc}
\end{equation}
where $P_\mathrm{in}$ is the on-chip pump tone power and $\Delta_1 = (\omega_0 - \delta\omega_0/2) - \omega_p$ is the effective detuning between pump tone and power-shifted cavity.
Lines in \textbf{a} show linear fits, from which we determine the shift per photon, i.e., the Kerr nonlinearity.
In \textbf{b}, the nonlinearity is plotted vs flux bias value with a sweetspot nonlinearity of $\chi \sim 2\pi\cdot 4\,$kHz.
We note, that depending on the pump frequency, we get quite significant variations in the extracted sweetspot nonlinearity with a range roughly between $1.5\,$kHz and $4\,$kHz, which we attribute to imprecise estimates of the intracavity photon number due to frequency dependent pump power arriving at the resonator input.
This can be explained by cable resonances and a frequency dependent transmission in the setup.
The values shown here are the largest we obtained.
The intrinsic linewidth itself is also photon-number dependent as discussed in the next paragraph.
This, however, does not impact significantly the calculation of $n_c$ by Eq.~(\ref{eqn:nc}) in our two-tone configuration, as $\Delta_1/\kappa \sim 10^2 - 10^3$.
\begin{figure}[h]
	\centerline {\includegraphics[trim={0cm 17.5cm 0cm 1cm},clip=True,scale=0.75]{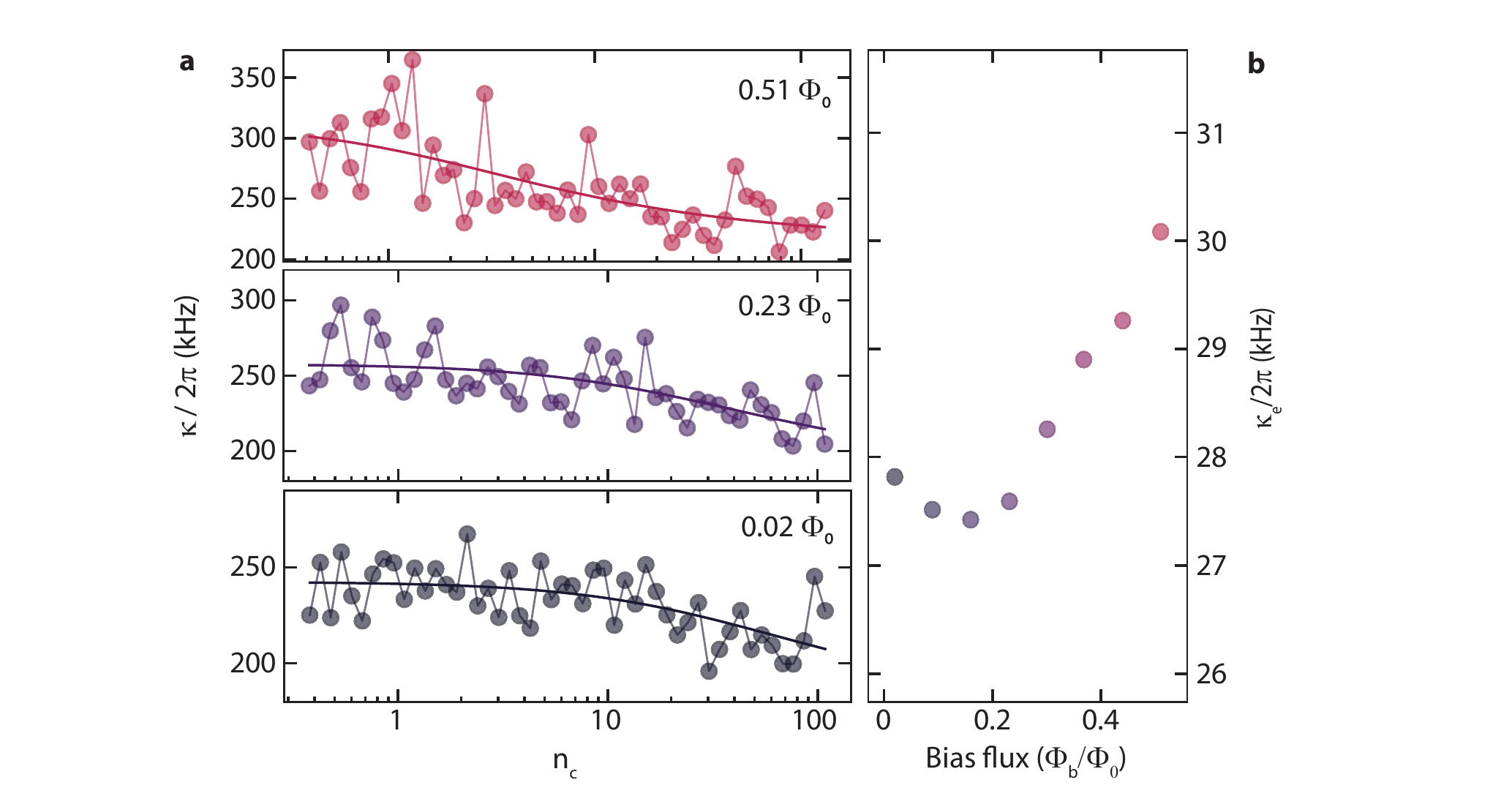}}
	\caption{\textbf{Characterizing the SQUID linewidth power dependence.} \textbf{a} Total SQUID cavity linewidth depending on intracavity photon number, shown for three different flux bias points. Circles are data, lines are fits using Eq.~(\ref{eqn:TLS}). In \textbf{b}, we show the values obtained for the external linewidth vs bias flux.}
	\label{fig:PowDepKappa}
\end{figure}
In Fig.~\ref{fig:PowDepKappa}\textbf{a}, we show the total SQUID cavity linewidth vs intracavity photon number, once again for different values of SQUID flux bias.
For better visibility, we restricted the shown data to three different flux bias values.
The scattering of the data points is cause by the measurement details of this measurement series (low probe power, high bandwidth, large frequency span), which leads to large fit uncertainties in the linewidth.
For all flux values, however, the linewidth follows a similar, decreasing trend with increasing photon number, which we interpret as indication for two-level system losses \cite{Hunklinger76, Pappas11}.
The power dependence for two-level system (TLS) losses is given by $\kappa_\mathrm{TLS}(n_c) = \kappa_\mathrm{TLS}/\sqrt{1 + \frac{n_c}{n_\mathrm{crit}}}$ with the critical photon number $n_\mathrm{crit}$ being a measure for the TLS saturation photon number.
The total linewidth therefore is given by
\begin{equation}
\kappa = \kappa_e + \kappa_1 + \frac{\kappa_\mathrm{TLS}}{\sqrt{1+\frac{n_c}{n_\mathrm{crit}}}}
\label{eqn:TLS}
\end{equation}
where $\kappa_1$ represents the internal losses not related to the power-dependent TLS losses and $\kappa_1 + \kappa_\mathrm{TLS}(n_c) = \kappa_i(n_c)$.
We fit the data with this Eq.~(\ref{eqn:TLS}) shown in Fig.~\ref{fig:PowDepKappa}\textbf{a} and find good agreement with the experimental observations.
For a cleaner dataset with less linewidth scattering, cf. Fig.~\ref{fig:PowerStrongCoupling}\textbf{b}.
The total linewidth slightly increases with flux, which could be either caused by an increased quasiparticle density or by a flux-noise induced linewidth broadening. 
In \textbf{c}, we plot the external linewidth $\kappa_e$ vs bias flux, and observe that is nearly independent of flux.

\section{Response functions and fitting routine}

\subsection{Ideal HF cavity and LF resonator response function}
Both, our HF SQUID cavity and the LF resonator, can be modeled as a parallel LC circuit capacitively coupled to a transmission line in a reflection geometry.
The $S_{11}$ response function of such a circuit is given by
\begin{equation}
S_{11} = 1- \frac{2\kappa_e}{\kappa_i+\kappa_e+2i\Delta}
\label{eq:Responsefunc}
\end{equation}
with detuning from the resonance frequency
\begin{eqnarray}
\Delta = \omega - \omega_0.
\end{eqnarray}
For the LF resonator, we get fully equivalently
\begin{equation}
S_{11} = 1- \frac{2\Gamma_e}{\Gamma_i+\Gamma_e+2i\Delta_0}
\label{eq:Responsefunc}
\end{equation}
with $\Delta_0 = \Omega - \Omega_0$.

\subsection{Real response function and fitting routine}

When analyzing the measured cavity response, we consider a frequency dependent complex-valued microwave background with amplitude and phase modulations originating from a variety of microwave components in our input and output lines and possible interfering signal paths.
Under this assumption, we model the modified cavity response with
\begin{eqnarray}
S_{11} = (\alpha_0 + \alpha_1\omega)\left(1-\frac{2\kappa_ee^{i\theta}}{\kappa_i+\kappa_e+2i\Delta}\right)e^{i(\beta_1\omega + \beta_0)}
\label{eq:fitS11}
\end{eqnarray}
where we consider a frequency dependent complex background
\begin{eqnarray}
S_{11} = (\alpha_0 + \alpha_1\omega)e^{i(\beta_1\omega + \beta_0)}
\label{eqn:Fit_BG}
\end{eqnarray}
and an additional rotation of the resonance circle due to the phase factor $e^{i\theta}$ \cite{}.
The first step in the fitting routine removes the cavity resonance part from the data curve and fits the remaining background with Eq.~(\ref{eqn:Fit_BG}).
After removing the background contribution from the full dataset by complex division, the resonator response is fitted using the ideal response function.
In the final step, the full function is re-fitted to the bare data using as starting parameters the individually obtained fit numbers from the first two steps.
From this final fit, we extract the final background fit parameters and remove the background of the full dataset by complex division.
Also, we correct for the additional rotation factor $e^{i\theta}$.
As result we obtain clean resonance curves as shown in Figs.~\ref{fig:LFfit} and \ref{fig:HFfit}.

\section{Parametrically coupled LC oscillators}

\subsection{Classical equations of motion}

We model the system for most experimental parts classically with the equations of motion for an oscillating magnetic flux $\Phi_\mathrm{LF}$ threading the SQUID loop, an analogue for the mechanical displacement in a typical optomechanical system, and for the SQUID cavity intracavity field amplitude $\alpha$
\begin{eqnarray}
\ddot{\Phi}_\mathrm{LF}  & = & - \Gamma_0\dot{\Phi}_\mathrm{LF} - \Omega_0^2\Phi_\mathrm{LF} + \frac{\hbar \gamma G|\alpha|^2}{C_\mathrm{LF}} + \sqrt{\Gamma_e}S_\mathrm{LF}\\
\dot{\alpha} & = & \left[i(\Delta + \gamma G\Phi_\mathrm{LF}) - \frac{\kappa}{2}\right]\alpha + \sqrt{\kappa_e}S_\mathrm{in}.
\end{eqnarray}
Here $C_\mathrm{LF}$ is the capacitance of the low frequency resonator, $\Delta = \omega_\mathrm{p} - \omega_0$ is the detuning of a pump tone from the cavity resonance frequency, $\kappa$ and $\Gamma_0$ are the total resonator linewidths and $\kappa_e$ and $\Gamma_e$ are the external linewidth of high and low-frequency circuit, respectively.
The quantities $S_\mathrm{in}$ and $S_\mathrm{LF}$ are the normalized input fields on the high-frequency and low-frequency input line, respectively.
The photon pressure (in optomechanical system radiation pressure force) from the high frequency cavity to the LF resonator is taken into account with the term $\hbar \gamma G|\alpha|^2/C_\mathrm{LF}$ with pull parameter $G$
\begin{equation}
G = -\frac{\partial\omega_0}{\partial \Phi}.
\end{equation}
The dimensionless parameter $\gamma = M/L_\mathrm{LF}$ describes the amount of the LF resonator flux coupling into the SQUID loop with the mutual inductance $M$ between the LF inductor $L_\mathrm{LF}$ and the SQUID loop inductance $L_l$.
Assuming that the intracavity field is high enough to only consider small deviations from the steady state solutions with $\Phi_\mathrm{LF} = \bar{\Phi}_\mathrm{LF} + \delta \Phi_\mathrm{LF}$ and $\alpha = \bar{\alpha} + \delta \alpha$ and no external driving of the low frequency resonator $S_\mathrm{LF}  = 0$, the equations of motion can be linearized as
\begin{eqnarray}
\delta\ddot{\Phi}_\mathrm{LF} & = & - \Gamma_0\delta \dot{\Phi}_\mathrm{LF} -\Omega_0^2\delta \Phi_\mathrm{LF} + \frac{\hbar \gamma G \bar{\alpha}} {C_\mathrm{LF}}(\delta \alpha + \delta\alpha^*)\\
\delta\dot{\alpha} & = & \left[ i\bar{\Delta} - \frac{\kappa}{2}\right]\delta \alpha + i\gamma G\bar{\alpha}\delta\Phi_\mathrm{LF} + \sqrt{\kappa_e}S_{p} 
\end{eqnarray}
In the above expressions, the detuning takes into account the shift from the equilibrium flux value, $\bar{\Phi}_\mathrm{LF}$ due to the radiation pressure force $\bar{\Delta} = \omega_\mathrm{p} - \omega_0 +G\bar{\Phi}_\mathrm{LF}$ and $\sqrt{\kappa_e}S_p$ with $S_p = S_0e^{-i(\omega_\mathrm{pr} - \omega_\mathrm{p})t}$ accounts for field fluctuations with the frequency $\omega_\mathrm{pr}$.
On a first look, it might seem surprising, that the interaction is related to a DC equilibrium flux in the LF resonator.
It corresponds to an additional, effective DC flux in the SQUID loop, generated by the HF currents in presence of an external SQUID flux bias.
As by the mutual inductance, the flux in the SQUID loop is part of the LF flux, the origin and interpretation of this DC flux becomes clear.
By solving the equations of motion we get the modified low-frequency resonator susceptibility
\begin{equation}
\chi_0^{\textrm{eff}} = \frac{1}{\Omega_0^2 - \Omega^2 - i\Omega\Gamma_0 - 2i\Omega_0 g^2\left[\chi_c(\Omega) - \chi_c^*(-\Omega)\right]}
\label{eqn:chi0eff}
\end{equation}
with the multi-photon-coupling rate
\begin{equation}
g = \bar{\alpha} G \Phi_\mathrm{zpf} = \sqrt{n_c} G \Phi_\mathrm{zpf}
\end{equation}
and the SQUID cavity susceptibility
\begin{eqnarray}
\chi_c = \frac{1}{\frac{\kappa}{2} - i(\bar{\Delta} + \Omega)}.
\end{eqnarray}
From here on and in the main paper, we just use $\Delta = \bar{\Delta}$ and always refer to the detuning as the detuning from the pump-shifted cavity resonance frequency.
For the SQUID cavity response, we get
\begin{equation}
S_{11} = 1-\kappa_e\chi_c\left[1+2i\Omega_0g^2\chi_c\chi_0^\mathrm{eff}\right].
\label{eqn:OMITdata}
\end{equation}

\subsection{Zero-point flux fluctuations and single-photon coupling rate} 

\begin{figure}[h]
	\centerline {\includegraphics[trim={1cm 20cm 1cm 0cm},clip=True,scale=0.65]{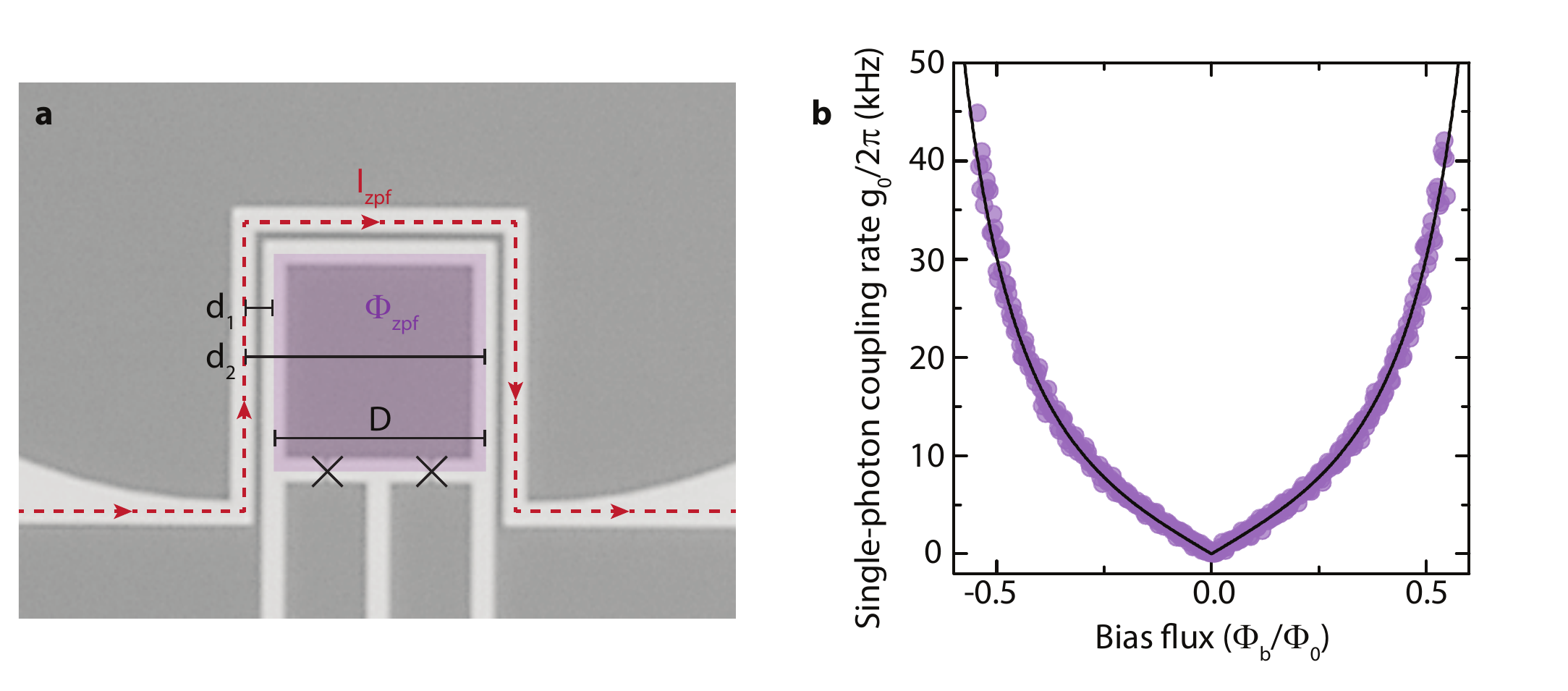}}
	\caption{\textbf{Calculation of the single-photon coupling rate $g_0$.} \textbf{a} Optical image showing the SQUID loop and the low frequency inductor wire. Arrows represent the geometrical distances $d_1$, $d_2$ and $D$, used for the calculation of the LF resonator induced zero-point SQUID flux $\Phi_\mathrm{zpf}$. The crosses indicate the positions of the Josephson junctions. \textbf{b} shows the calculated single-photon coupling rate vs SQUID bias flux. The points are calculated from the measured SQUID resonance frequency, the line from the fit to the flux arch.}
	\label{fig:ZPflux}
\end{figure}

The zero-point current of the low frequency resonator is given by $I_\textrm{zpf} = 21\,$nA.
The zero-point flux fluctuations generated in the SQUID loop can therefore be calculated by estimating the flux $\Phi_\mathrm{zpf} = MI_\mathrm{zpf}$ induced by $I_{\textrm{zpf}}$ in the loop, where $M$ denotes the mutual inductance between the LF wire and the SQUID loop.
Figure~\ref{fig:ZPflux} shows an optical image of the SQUID loop and the LF wire passing by three sides of the loop with a $500\,$nm wide gap between the $1\,\mu$m wide wires.
With the geometrical parameters of the configuration, $D=10\,\mu$m being the length of one loop side, $d_1 = 1\,\mu$m being the distance between the LF wire and the nearest SQUID loop wire, and $d_2 = 11\,\mu$m being the distance between the LF wire and the distant SQUID loop wire, the flux induced by $I_\mathrm{zpf}$ is calculated via
\begin{equation}
\Phi_\textrm{zpf} = 3 \frac{\mu_0}{2\pi}I_\textrm{zpf} D \ln \left( \frac {d_2}{d_1} \right) \approx 145\,\mu\Phi_0.
\label{eq:fluxzpf}
\end{equation}
For all geometrical distances $D$, $d_1$, and $d_2$, we considered the center of the corresponding wires.
This also gives directly an estimate for the mutual inductance $M = 14\,$pH.
The single-photon coupling rate is given by
\begin{equation}
g_0 = \frac{\partial \omega_0}{\partial\Phi} \Phi_\textrm{zpf}
\end{equation}
where $\Phi_\textrm{zpf}= 145\, \mu\Phi_0$ are the zero-point flux and the flux responsivity is extracted from the SQUID cavity flux dependence, cf. Fig.~\ref{fig:FluxArch}\textbf{c}.

Figure~\ref{fig:ZPflux} shows the calculated single-photon coupling rates of the device depending on the flux-bias point.
According to this calculation, the single-photon coupling rate can be tuned from $g_0 = 0\,$Hz, when the cavity is operated at the sweetspot, to $g_0 \approx 2\pi\cdot 40\,$kHz for the largest flux bias values.

\subsection{Dynamical photon-pressure backaction}

The term $\Sigma = -2i\Omega_0g^2[\chi_c(\Omega) - \chi_c^*(-\Omega)]$ in the modified LF resonator susceptibility $\chi_0^\mathrm{eff}$ can be understood as a SQUID cavity field induced term modifying the damping and the resonance frequency of the LF resonator.
This becomes apparent, when we assume the high-$Q_0$ limit, where the susceptibility for a red-sideband pump tone with $\Delta\approx-\Omega_0$ and $\Omega \approx +\Omega_0$ can be approximated as
\begin{equation}
\chi_0^\mathrm{eff} = \frac{1}{2\Omega_0}\frac{1}{(\Omega_m - \Omega_0) - i\frac{\Gamma_0}{2} - ig^2[\chi_c(\Omega) - \chi_c^*(-\Omega)]}
\end{equation}

By rewriting $\Sigma' = -ig^2[\chi_c(\Omega) - \chi_c^*(-\Omega)]$ as $\Sigma' = \delta\Omega_0  - i\delta\Gamma_0/2$  and independently analyze the real and imaginary part, we can write the change in frequency $\delta\Omega_0$ (photon-pressure frequency shift) and the additional damping term $\delta\Gamma_0$ (photon-pressure damping), arising from the modified susceptibility as
\begin{eqnarray}
\delta\Omega_0 & = & g^2 \left[ \frac{\Delta + \Omega_0}{\frac{\kappa^2}{4} + (\Delta + \Omega_0)^2} + \frac{\Delta - \Omega_0}{\frac{\kappa^2}{4} + (\Delta - \Omega_0)^2} \right]\\
\label{eq:spring}
\delta\Gamma_0 & = & g^2\kappa \left[ \frac{1}{\frac{\kappa^2}{4} + (\Delta + \Omega_0)^2} - \frac{1}{\frac{\kappa^2}{4} + (\Delta - \Omega_0)^2} \right].
\label{eq:damping}
\end{eqnarray}
For a blue-detuned pump field we find essentially the same expressions with a sign change for the photon-pressure damping $\delta\Gamma_0^\mathrm{blue} = -\delta\Gamma_0^\mathrm{red}$.
In order to observe the dynamical photon-pressure backaction between the two circuits, we flux bias the SQUID cavity with $\Phi_b \approx 0.14\Phi_0$ and stepwise sweep a pump tone through both, the red sideband $\Delta = - \Omega_0 + \delta_r$, and the blue sideband $\Delta = +\Omega_0 + \delta_b$, while the low-frequency resonator is scanned with a weak probe tone $\Omega_\mathrm{pr} = \Omega_0 + \Delta_0$.
The variables $\delta_r$, $\delta_b$ and $\Delta_0$ denote small detunings from the red sideband, the blue sideband and the LF resonance frequency, respectively.
As the device is deep in the resolved sideband regime with $\kappa/\Omega_0 \sim 1000$, we can approximate the expressions for the optical damping and the optical spring as
\begin{eqnarray}
\delta\Omega_0 & = & 4g^2\frac{\delta_i}{\kappa^2 + 4\delta_i^2}\\
\delta\Gamma_0 & = & \pm 4g^2\frac{\kappa}{\kappa^2 + 4\delta_i^2}
\label{eqn:DBfits}
\end{eqnarray}
with $i = b, r$ and $+$ for the red sideband and $-$ for the blue sideband.
From fits to the LF resonator response, we extract $\delta\Omega_0$ and $\delta\Gamma_0$ as shown and dicussed in the main paper Fig.~2.
The two expressions Eqs.~(\ref{eqn:DBfits}) are only strictly valid for constant $\kappa$ far away from the strong coupling regime $g \ll (\kappa-\Gamma_0)/4 \sim 2\pi\cdot 50\,$kHz.
When the strong-coupling regime is approached, the LF resonator linewidth is not a linear function of power anymore and the SQUID cavity linewidth starts to decrease \cite{Peterson19}.
The weak-coupling condition condition is not strictly fulfilled anymore in our experiment for cooperativites $\mathcal{C} \sim 1$ and hence we expect deviations.
When we use a power-dependent effective cavity linewidth $\tilde{\kappa}$ for these equations instead of the bare linewidth, however, we can describe the data perfectly with these expressions.

\subsubsection*{Dynamical backaction - Additional data}
\begin{figure}[h]
	\centerline {\includegraphics[trim={1cm 5cm 1cm 0cm},clip=True,scale=0.6]{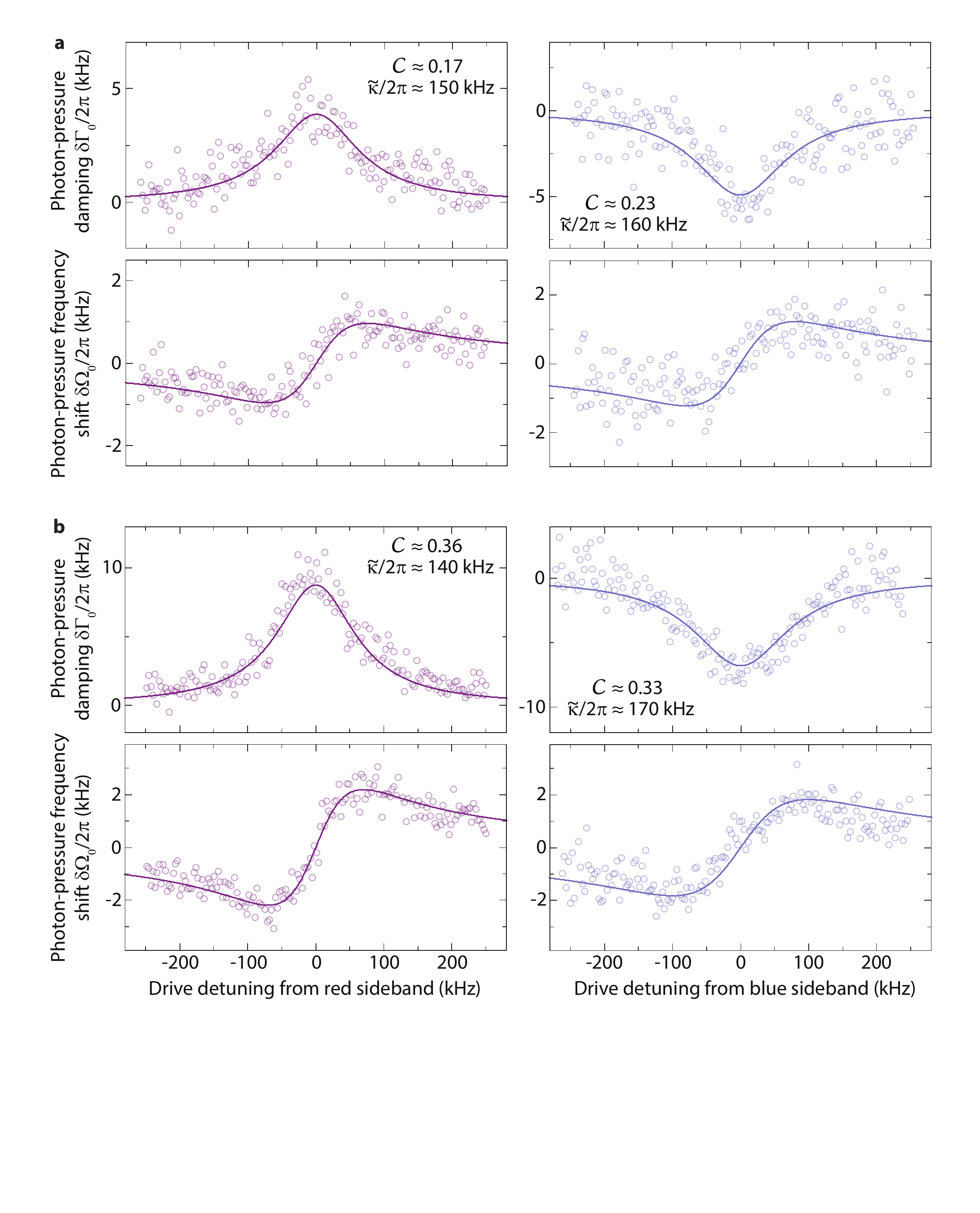}}
	\caption{\textbf{Dynamical backaction for red- and blue-detuned pump tone, respectively, in the low cooperativity regime.} \textbf{a} and \textbf{b} show the extracted photon-pressure damping and photon-pressure frequency shift for a pump on the red sideband (left) and for a pump on the blue sideband (right). \textbf{a} is for approximately 15 intracavity photons and \textbf{b} for 30. The flux bias point was $\Phi_b/\Phi_0 \approx 0.14$. Circles are data, lines are theoretical curves with Eqs.~(\ref{eqn:DBfits}) and the parameters given in the panels. On the red sideband, the effective cavity linewidth $\tilde{\kappa}$ needed to describe the data with the expressions (\ref{eqn:DBfits}) decreases with pump strength/cooperativity, on the blue sideband it decreases. This indicates that it is not an effect of nonlinear cavity losses, but deviations from the used approximations due to the similarity of $\kappa, g$ and $\Gamma_0$ here, i.e., due to approaching the strong-coupling regime already for $C \sim 1$ \cite{Peterson19}.}
	\label{fig:DB_Add}
\end{figure}

Here, we show additional data for dynamical photon-pressure backaction in Fig.~\ref{fig:DB_Add}, taken for two lower pump powers than used in the main paper data. The pump powers on the blue and on the red sideband were adjusted to give similar cooperativities.

\subsection{The SQUID cavity response function for a red-sideband pump}

With the modified low-frequency resonator susceptibility $\chi_0^\mathrm{eff}$ we can write the SQUID cavity response as 
\begin{equation}
S_{11} = 1-\kappa_e\chi_c\left[1+2i\Omega_0g^2\chi_c\chi_0^\mathrm{eff}\right].
\label{eqn:OMIT}
\end{equation}

\subsubsection*{Extracting cavity and coupling parameters from data}

\begin{figure}[h!]
	\centerline {\includegraphics[trim={1cm 5.3cm 2cm 0.1cm},clip=True,scale=0.74]{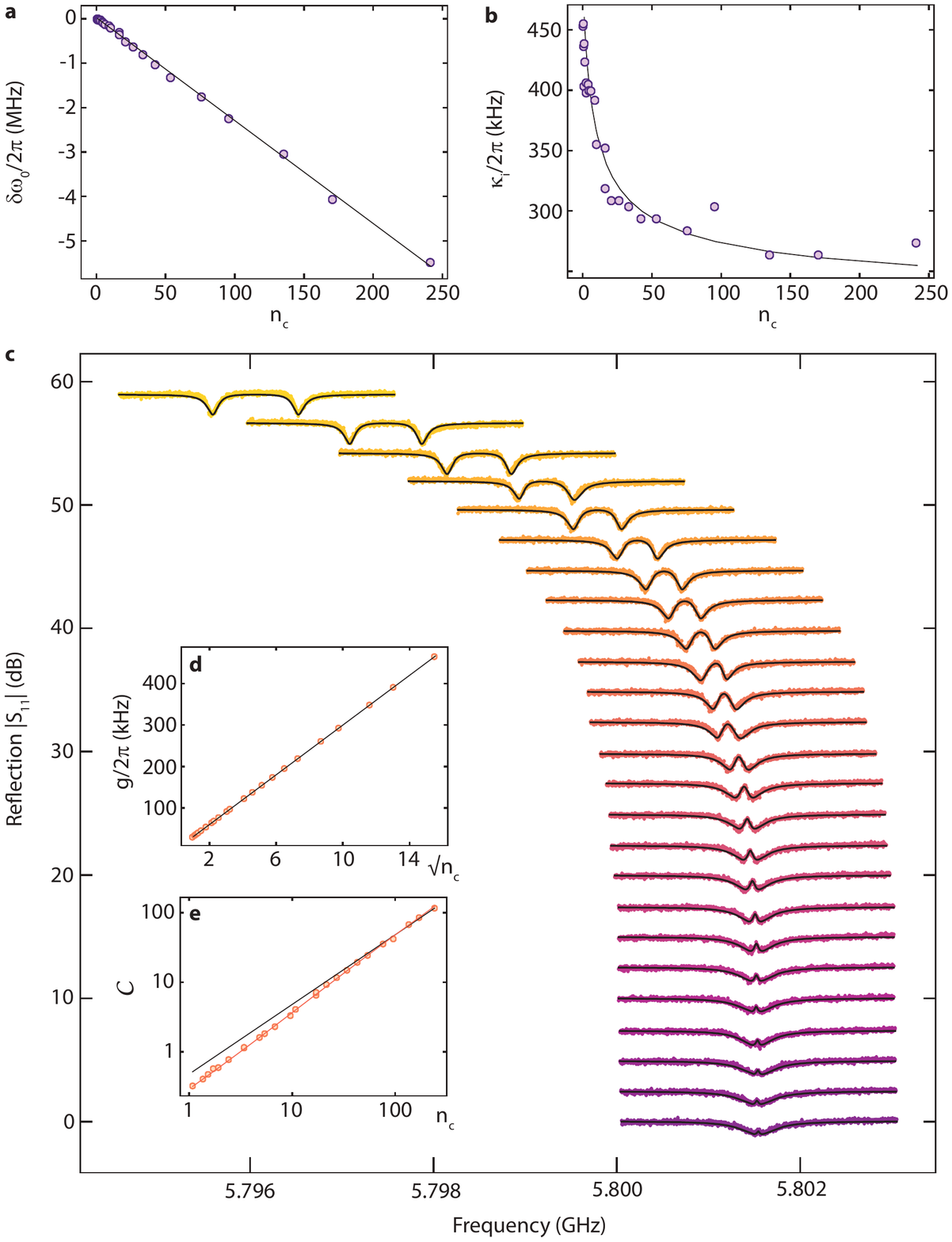}}
	\caption{\textsf{\textbf{Transition to the strong-coupling regime by increasing the intracavity photon number $n_c$.} \textbf{a} Resonance frequency shift from the bare cavity resonance $\delta\omega_0 = \omega_0'-\omega_0$ with $n_c$, where $\omega_0$ is the resonance frequency at the lowest power. The points are data and the black line is a linear fit. \textbf{b} Internal linewidth $\kappa_i$ of the HF cavity depending on $n_c$. The points are the extracted values from adjusting the theoretical curves to the data presented in \textbf{c}. The black line is a fit curve based on Eq.~\ref{eqn:TLS}. \textbf{c} Observation of the transition from photon pressure-induced transparency to the strong-coupling regime by increasing the power of a pump tone on the red sideband of the SQUID cavity $\omega_\mathrm{p} = \omega_0 - \Omega_0$. The colored points are the measured data and the black lines are theoretical curves. Each of the curves was upshifted by 2.5 dB for better visibility with the lowest curve being unshifted. \textbf{d} Linear increase of the coupling rate $g$ with $\sqrt{n_c}$, where the points are the extracted values from the data and the line is a fit. \textbf{e} Cooperativity of the system vs intracavity photon number, plotted on a logarithmic scale. The black line is a linear fit (valid for constant $\kappa$) and the orange line is the cooperativity calculated with the internal linewidth of the cavity based on the fit curve presented in \textbf{b}.}}
	\label{fig:PowerStrongCoupling}
\end{figure}

To model the data of the response in presence of a red sideband pump, we use Eq.~(\ref{eqn:OMIT}) and adjust the parameters $\omega_0$, $\kappa$, $\Omega_0$ and $g$ for fixed $\Gamma_0 = 2\pi\cdot 22\,$kHz and $\kappa_e = 2\pi\cdot28\,$kHz.
Corresponding lines ae shown in the main paper Fig.~3\textbf{d} and Fig.~\ref{fig:PowerStrongCoupling}\textbf{c}.
With the extracted numbers, we calculate the cooperativity
\begin{equation}
\mathcal{C} = \frac{4g^2}{\kappa\Gamma_0}
\end{equation}
and estimate the SQUID cavity photon number via
\begin{equation}
n_c = \frac{4P_\mathrm{in}}{\hbar\omega_\mathrm{p}}\frac{\kappa_e}{\kappa^2 + 4\Delta^2}
\end{equation}
where $P_\mathrm{in}$ is the on-chip power calculated from the generator output power and the estimated input attenuation of $-61\,$dB.
From those numbers, we calculate the single-photon coupling rate $g_0 = \frac{g}{\sqrt{n_c}}$. 

\subsection{The strong-coupling regime}
When increasing the multi-photon optomechanical coupling rate $g = \sqrt{n_c}g_0$ to a point where $g>(\kappa-\Gamma_0)/4$, the system enters the strong-coupling regime \cite{Peterson19}, where the driven high-frequency mode and the low-frequency mode hybridize, forming two new modes split by $2g$, an effect also known as normal-mode splitting.
When the high-frequency cavity is pumped exactly on the red-sideband  $\Delta = \omega_\mathrm{p} - \omega_0 = -\Omega_0$, the two new formed hybrid excitation modes of the system have (complex) eigenfrequencies given by \cite{Aspelmeyer14}
\begin{eqnarray}
\omega_{\pm} = \Omega_0 - i\frac{\kappa+\Gamma_0}{4} \pm \sqrt{g^2 - \left(\frac{\kappa-\Gamma_0}{4} \right)^2},
\label{eqn:NMS}
\end{eqnarray}

with $\omega_+ - \omega_- = 2g$ for $g^2 \gg (\kappa - \Gamma_0)^2/16$.
The real part of $\omega_{\pm}$ describes the resonance frequencies of the two hybridized modes and the imaginary part is half the linewidth.
In the strong-coupling regime $g^2 \gg (\kappa - \Gamma_0)^2/16$, we get
\begin{equation}
\omega_\pm = \Omega_0 \pm g - i\frac{\kappa+\Gamma_0}{4}
\end{equation}
showing that each of the new modes has a linewidth $\Gamma_\pm = (\kappa + \Gamma_0)/2$.
We also note here, that the formal definition of the strong-coupling regime here with $g>(\kappa -\Gamma_0)/4$, corresponding to a sudden transition from split damping rates to split eigenfrequencies in Eq.~(\ref{eqn:NMS}), is different from the standard definition, where the mode splitting $2g$ must exceed the hybridized mode linewidths $(\kappa+\Gamma_0)/2$.
This would correspond to $g>(\kappa+\Gamma_0)/4$.
Our device, however, reaches the strong-coupling regime unambiguously for either of the definitions.
\subsection{Transition to the strong-coupling regime by increasing $n_c$}

In Fig.~3 of the main paper, we show the transition to the strong-coupling regime by increasing the single-photon coupling strength $g_0$ through the flux responsivity $\partial\omega_0/\partial\Phi$, while keeping the red-sideband pump strength constant.
A similar transition to the strong-coupling regime can also be accomplished by the common approach of enhancing the coupling strength of the system by increasing the number of photons in the cavity \cite{Teufel11}.
The experimental setup corresponds to what is described in Fig.~3\textbf{a} of the main paper, where a pump tone was set to the red sideband of the high-frequency cavity $\omega_\mathrm{p} = \omega_0 - \Omega_0$, and a probe tone was scanning through the cavity resonance $\omega_\mathrm{pr} \approx \omega_0$.
The presented measurements in this section were performed at $\Phi_b \approx 0.5\Phi_0$ and the external measurement configuration is presented in detail in Fig.~\ref{fig:Setup}\textbf{c}.
The measurement was repeated for increasing values of drive power.
Due to the residual SQUID cavity Kerr-nonlinearity, we have to adjust the pump frequency for each power to set the pump onto the actual red sideband.
The cavity resonance frequency and linewidth vs photon number is shown in Fig.~\ref{fig:PowerStrongCoupling}\textbf{b} and \textbf{c}.
Note, that this experiment was done during a later cooldown compared to most other experiments and the device parameters (linewidths and resonance frequencies) seem to have somewhat shifted in between the two cooldowns.
In  Fig.~\ref{fig:PowerStrongCoupling}\textbf{c}, we plot the SQUID cavity response vs probe frequency for different pump powers, with increasing power from bottom to top.
For the lowest power, the cavity resonance is almost unperturbed besides a very little interference peak in its center.
With increasing pump power and therefore increasing intracavity photon number $n_c$, the PPIT interference effect of the low-frequency resonator in the high-frequency cavity response gets enhanced until for the highest powers a clear normal-mode splitting with up to $g/\pi = 950\,$kHz occurs.
We model the data with Eq.~(\ref{eqn:OMITdata}). 
For each of the theoretical curves, the parameters $\kappa_e$, $\Gamma_0$ and $g_0$ were fixed at $2\pi\cdot28\,$kHz, $2\pi\cdot25\,$kHz and $2\pi\cdot30\,$kHz, respectively.
To get agreement with the experimental data, we varied cooperativity, cavity frequency shift, LF resonator resonance frequency and internal linewidth of the SQUID cavity.
The values for $\delta\omega_0$, $\kappa_i$ and $\mathcal{C}$ as extracted are plotted in Fig.~\ref{fig:PowerStrongCoupling}\textbf{a}, \textbf{b} and \textbf{e}, respectively.
From the combination, we also determined the multi-photon coupling rate $g$ as plotted in \textbf{d}. 
The observed behaviour of the internal linewidth with intracavity photon number follows the prediction for two-level system (TLS) losses as given by Eq.~(\ref{eqn:TLS}).
Furthermore, the extracted cooperativities in Fig.~\ref{fig:PowerStrongCoupling}\textbf{e} do not show a stricly linear behaviour, cf. the linear fit (black line).
Including the internal linewidth decrease with photon number due to TLS losses, as shown as orange line, gives excellent agreement with the data points.

\subsection{Detecting normal-mode splitting in the LF resonator response}

\begin{figure}[h]
	\centerline {\includegraphics[trim={1cm 16.3cm 2cm 0.6cm},clip=True,scale=0.6]{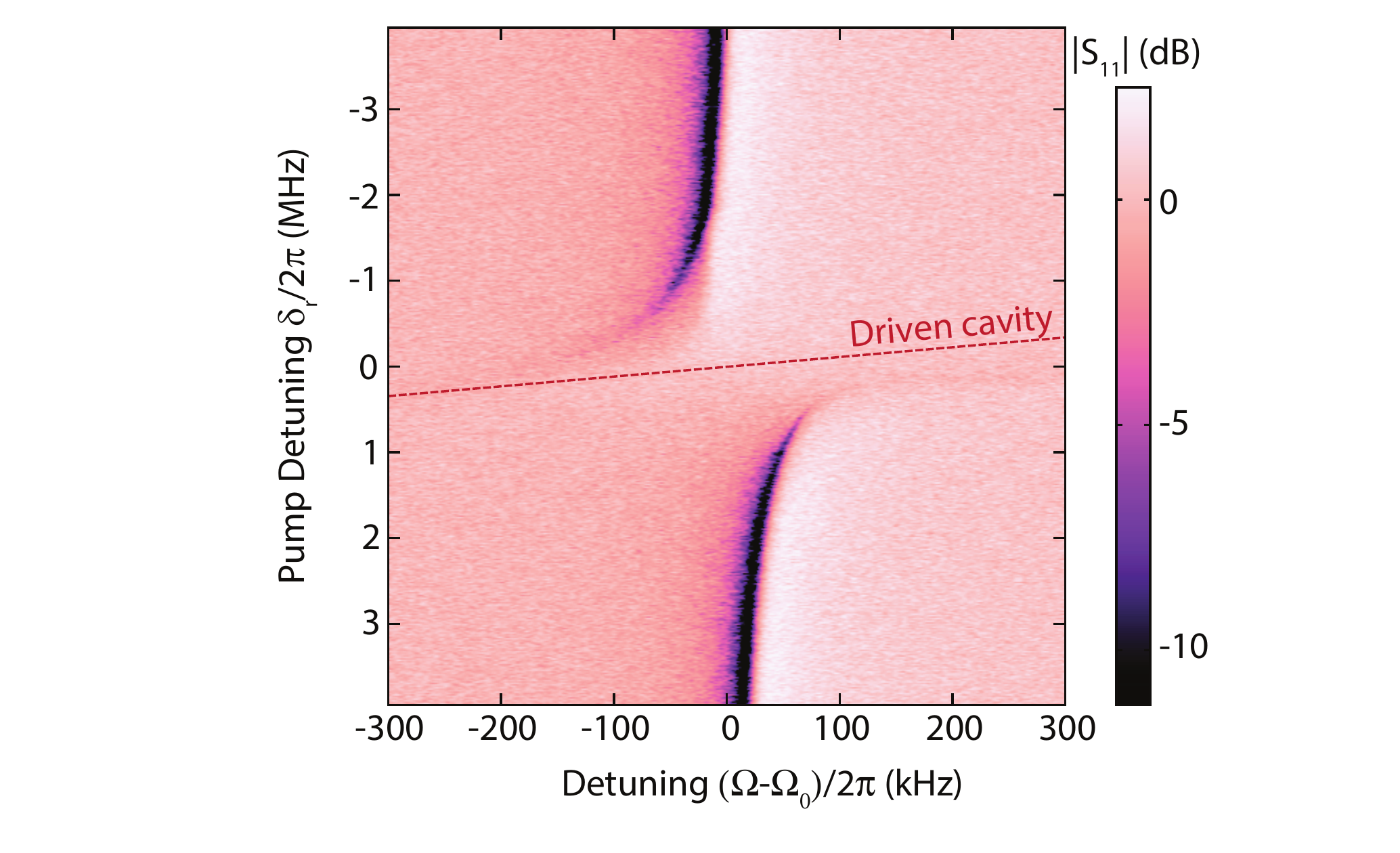}}
	\caption{\textsf{\textbf{Normal-mode splitting of the LF resonator response.} When measuring the response of the LF resonator with the parameters as in main paper Fig.~3\textbf{e}, case D, i.e., with a red-detuned SQUID cavity drive, $n_c\sim 70$ intracavity photons and $\Phi_b/\Phi_0 \sim 0.5$, we observe normal-mode splitting in the direct response of the LF resonator. The red dashed line indicates the resonance frequency of the SQUID cavity mode with respect to the pump tone.}}
	\label{fig:LFSC}
\end{figure}
The normal-mode splitting in the strong-coupling regime can not only be detected in the SQUID cavity response, but also in the LF resonator response.
When probing the LF resonator under the experimental conditions of the main paper Fig.~3\textbf{e}, case D, we observe the LF response as shown in Fig.~\ref{fig:LFSC} with a pronounced normal-mode splitting.
The dashed red line shows the SQUID cavity resonance frequency with respect to the moving red-sideband pump.

\section{Detection and amplification of thermal noise}

\subsection{Power spectral density for a blue-sideband drive}

Following the routine given in the Supplementary Material of Ref.~\cite{Teufel11}, we derive the cavity field amplitude for a blue sideband drive to be given by
\begin{equation}
\hat{a} = \frac{-ig\chi_c\bar{\chi}_0\sqrt{\Gamma_0}\hat{S}_0^\dagger - \chi_c\sqrt{\kappa_e}\hat{S}_c}{1-g^2\chi_c\bar{\chi}_0}
\end{equation}
with the susceptibilities
\begin{eqnarray}
\chi_c & = & \frac{1}{\frac{\kappa}{2} + i(\Omega + \Delta)}\\
\bar{\chi}_0 & = & \frac{1}{\frac{\Gamma_0}{2} - i(-\Omega - \Omega_0)}.
\end{eqnarray}
As usual, $\Delta = \omega_\mathrm{p} - \omega_0 \sim \Omega_0$ describes the detuning between cavity resonance and the blue sideband pump here, and $\Omega \sim - \Omega_0$ is the frequency where we measure and calculate the output field.
The operators $\hat{S}_0$ and $\hat{S}_c$ denote the low frequency resonator and SQUID cavity noise input fields and follow $\langle\hat{S}_i\hat{S}_i^\dagger\rangle = n_i+1$ and $\langle\hat{S}_i^\dagger\hat{S}_i\rangle = n_i$.

\subsubsection*{Added noise}

According to Ref. \cite{Teufel11}, the effective number of added noise photons is given by

\begin{equation}
n_\mathrm{add}' = \frac{n_\mathrm{add}}{\eta}+\left(\frac{1-\eta}{\eta}\right)\frac{1}{2}
\end{equation}

where $n_\mathrm{add}$ is the actual number of photons added by the HEMT amplifier in our case, $n_\mathrm{add}\sim 20$, and $\eta \sim 0.7$ accounts for losses of the cavity output field on its way to the HEMT. 
Thus, we estimate the effective total number of added photons to be $n_\mathrm{add}' \approx 29$.

\subsubsection*{The total power spectral density}

For the power spectral density at frequency $\omega = \omega_\mathrm{p} + \Omega$ of the SQUID cavity with a drive around one of the sidebands, we get with this

\begin{equation}
\frac{S(\omega)}{\hbar\omega} = \frac{1}{2} + n_\mathrm{add}' + \frac{\kappa_e g^2 |\bar{\chi}_0|^2 |\chi_c|^2 \Gamma_0 \left(n_\mathrm{LF}+1\right) + \kappa_e |\chi_c|^2\kappa n_c}{|1 - g^2\chi_c\bar{\chi}_0|^2}
\end{equation}

The thermal mode occupations follow a Bose distribution, i.e.,

\begin{eqnarray}
n_c & = & \frac{1}{e^{\frac{\hbar\omega_0}{k_\mathrm{B}T_c}}-1}\\
n_\mathrm{LF} & = & \frac{1}{e^{\frac{\hbar\Omega_0}{k_\mathrm{B}T_0}}-1}
\end{eqnarray}

and we assume the added noise photons $n_\mathrm{add} = k_\mathrm{B}T_\mathrm{HEMT}/\hbar\omega$ to be given by the noise temperature of the cryogenic HEMT amplifier with $T_\mathrm{HEMT} = 5.5\,$K.
We note that we decided to choose possibly different temperatures for the low-frequency and the high-frequency thermal distribution, as the two resonators are isolated differently from the noise of their environment, e.g. the input/output cabling.
Assuming a negligible SQUID cavity occupancy $n_c \sim 0$, we can rewrite the power spectral density as

\begin{equation}
\frac{S(\omega)}{\hbar\omega} = \frac{1}{2} + n_\mathrm{add}' + \frac{16\kappa_e g^2 \Gamma_0}{(\Gamma_0'^2 + 4\Delta'^2)(\kappa^2 + 4\delta_b^2)}\left(n_\mathrm{LF} + 1\right)
\end{equation}

where $\Gamma_0' = \Gamma_0 -\delta\Gamma_0$ includes the effect of photon-pressure damping, the detuning $\Delta' = -\Omega - (\Omega_0 + \delta\Omega_0)$ includes the photon-pressure frequency shift, and $\delta_b = \omega_\mathrm{p} - (\omega_0 + \Omega_0)$ takes into account possible detunings of the pump tone from the blue cavity sideband.

\subsubsection*{PSD with pump exactly on the blue sideband}

When the blue sideband pump is exactly at $\omega_\mathrm{p} = \omega_0 + \Omega_0$, i.e., $\delta_b = 0$, the photon-pressure frequency shift vanishes $\Delta' = \Delta$ and we get

\begin{equation}
\frac{S(\omega)}{\hbar\omega} = \frac{1}{2} + n_\mathrm{add}' + 4\frac{\kappa_e}{\kappa}\mathcal{C}\frac{\Gamma_0^2}{\Gamma_0'^2 + 4\Delta^2}(n_\mathrm{LF}+1)
\end{equation}

which describes a Lorentzian sitting on top of a background with $1/2 + n_\mathrm{add}'$ photons.
With the power spectral density of the thermal current fluctuations in the LF resonator

\begin{equation}
S_{I}(-\Omega) = \frac{8\Gamma_0}{\Gamma_0'^2 + 4\Delta^2}I_\mathrm{zpf}^2 (n_\mathrm{LF}+1)
\end{equation}

and the corresponding flux fluctuations coupling into the SQUID loop

\begin{equation}
S_{\Phi}(-\Omega) = \frac{8\Gamma_0}{\Gamma_0'^2 + 4\Delta^2}M^2\Phi_\mathrm{zpf}^2 (n_\mathrm{LF}+1)
\end{equation}

where $M$ is the mutual inductance between the LF inductor and the SQUID loop, we can write

\begin{equation}
\frac{S(\omega)}{\hbar\omega} = \frac{1}{2} + n_\mathrm{add}' + \frac{\mathcal{C}}{2}\frac{\kappa_e}{\kappa}\frac{\Gamma_0}{ I_\mathrm{zpf}^2}S_I(-\Omega).
\end{equation}

\subsection{Extraction of thermal photon numbers from data}

For the extraction of the thermal photon numbers, we first determine $\mathcal{C}$, $\Gamma_0'$, $\kappa$ and $\kappa_e$ from a measurement of photon-pressure induced absorption, cf. main paper Fig.~4\textbf{a}.
In addition, we use the calculated zero-point fluctuations $I_\mathrm{zpf} = 21\,$nA and the added noise photons $n_\mathrm{add}' = 29$.
With the background noise amplitude $S_b = G_m\hbar\omega\left(\frac{1}{2} + n_\mathrm{add}'\right)$ with $G_m$ being the total gain of the amplifier detection line, determined from a Lorentzian fit of the PSD data $S_V$, we calculate
\begin{equation}
S_I(-\Omega) = \left[\frac{S_V}{S_b}-1\right]\left[\frac{1}{2} + n_\mathrm{add}'\right]\frac{2\kappa}{\mathcal{C}\kappa_e\Gamma_0}I_\mathrm{zpf}^2.
\end{equation}
To extract the thermal number of photons in the LF resonator, we use the herewith calculated $S_I(-\Omega)$ and consider the amplitude at resonance
\begin{equation}
S_{I0} = 8\frac{\Gamma_0}{\Gamma_0'^2}I_\mathrm{zpf}^2(n_\mathrm{LF}+1)
\end{equation}
which allows to calculate the thermal photon number according to
\begin{equation}
n_\mathrm{LF} = S_{I0}\frac{\Gamma_0'^2}{8\Gamma_0 I_\mathrm{zpf}^2}-1.
\end{equation}

\subsection*{Supplementary References}

\end{document}